\documentclass[11pt]{article}
\usepackage{amssymb,amsfonts,amscd,cite,amsmath,graphicx,epsf}
\usepackage{epstopdf}
\usepackage{psfrag}
\usepackage{enumerate}
\usepackage{cases}
\usepackage{bbm}

\usepackage[totalwidth=161mm,totalheight=245mm,centering]{geometry}

\newcommand\blank[1]{}

\newcommand{\fract}[2]{{\textstyle\frac{#1}{#2}}}

\newcommand{\ep}{\epsilon}

\newcommand\ZZ{{\mathbb Z}}
\newcommand\RR{{\mathbb R}}

\newcommand\phup{^{\phantom o}}

\newcommand{\CM}{{\cal M}}

\newcommand\beq{\begin{equation}}
\newcommand\eeq{\end{equation}}
\newcommand\beqa{\begin{eqnarray}}
\newcommand\eeqa{\end{eqnarray}}
\newcommand\nn{\nonumber}
\newcommand\half{{\textstyle\frac{1}{2}}}

\newcommand{\resection}[1]{\setcounter{equation}{0}\section{#1}}
\newcommand{\appsection}[1]{\addtocounter{section}{1}
\setcounter{equation}{0} \section*{Appendix \Alph{section}~~#1}}

\newcommand{\te}{\theta}

\newcommand{\tbi}{\theta_{b1}}
\newcommand{\tbii}{\theta_{b2}}
\newcommand{\htbi}{\hat\theta_{b1}}
\newcommand{\htbii}{\hat\theta_{b2}}
\newcommand{\tz}{\theta_0}

\newcommand{\gb}{g\phup_b}
\newcommand{\go}{g\phup_{0}}
\newcommand{\goA}{g\phup_{A}}
\newcommand{\goB}{g\phup_{B}}
\newcommand{\gi}{g\phup_{b1}}
\newcommand{\gii}{g\phup_{b2}}
\newcommand{\giii}{g\phup_{b3}}
\newcommand\larr{\!\!\!\xleftarrow{~\quad}\!\!\!}
\newcommand\rarr{\!\!\!\xrightarrow{~\quad}\!\!\!}
\begin{document}

\begin{titlepage} 
\vskip 0.5cm
\begin{flushright}
DCPT--10/35  \\
August 2010
\end{flushright}
\vskip 1.8cm
\begin{center}
{\Large\bf Exact g-function flows from the staircase model
} \\[5pt]
\end{center}
\vskip 0.8cm

\centerline{Patrick Dorey$^1$,
Roberto Tateo$^2$ and Ruth Wilbourne$^1$}
\vskip 0.9cm
\centerline{${}^{1}$\sl\small Department of Mathematical Sciences,
Durham University,}
\centerline{\sl\small South Road, Durham DH1 3LE, UK}
\vskip 0.3cm 
\centerline{${}^{2}$\sl\small Dip.\ di Fisica Teorica
and INFN, Universit\`a di Torino,} 
\centerline{\sl\small Via P.\ Giuria 1, 10125 Torino, Italy}
\vskip .35cm
\centerline{\small
{\tt p.e.dorey@durham.ac.uk}, {\tt tateo@to.infn.it},}
\centerline{\small {\tt r.m.wilbourne@durham.ac.uk}}
\vskip 1.9cm
\begin{abstract}
\noindent
Equations are found for exact $g$-functions corresponding to
integrable bulk and boundary flows between 
successive unitary $c<1$ minimal conformal field
theories in two dimensions, confirming and extending previous
perturbative results.
These equations are obtained via an embedding of the 
flows into a boundary version of Al.\,Zamolodchikov's 
staircase model.
\end{abstract}
\end{titlepage}
\setcounter{footnote}{0}

\resection{Introduction}
The exact $g$-function \cite{Dorey:2004xk,Dorey:2005ak} is a powerful 
tool for the study of integrable boundary flows, allowing the 
results of, 
for example, \cite{Tsvelick:1985,Affleck:1991tk} to be extended
to situations where the bulk is not critical.
The initial proposals of \cite{Dorey:2004xk} were restricted
to cases where the bulk theory possessed only massive excitations, and
their scattering was diagonal both in the bulk and at the boundary.
Recently, in \cite{Dorey:2009vg},
equations were introduced to describe the exact $g$-function
for the simplest case were massless bulk degrees of freedom
persist even in the far infrared, namely the flow between the
tricritical and critical Ising models. As mentioned in 
\cite{Dorey:2009vg},
these `massless' $g$-function
flow equations can be obtained from a consideration
of the so-called staircase models, and as a result are naturally embedded
in a much richer set of flows linking the boundary behaviours of
all of the unitary $c<1$ minimal 
models of conformal field theory. In this paper we will provide
some more details of this larger pattern, and in the process propose
equations to describe $g$-function flows between
all neighbouring pairs of unitary minimal models. For all of
these cases beyond the first (which is the already-discussed flow
between tricritical and critical Ising models), bulk scattering is
non-diagonal. However, the $g$-function equations have a simple form
which naturally generalises previously-seen 
structures\footnote{As we were writing this paper, an 
alternative
derivation of the diagonal $g$-function flow equations of 
\cite{Dorey:2004xk,Dorey:2005ak,Dorey:2009vg}, was presented, 
in \cite{Pozsgay:2010tv} (see also
\cite{Woynarovich:2010wt}).  It will be worthwhile, and appears in 
most respects to be straightforward, to generalise the approach 
of \cite{Pozsgay:2010tv} to cover our 
new equations, but as we feel the staircase
aspect is of independent interest we have decided to leave this
point for the time being. We should also mention that boundary
$g$-functions in the staircase model were previously discussed,
amongst other things, in
\cite{Lesage:1998qf}, but since that paper predated the exact
$g$-function results of \cite{Dorey:2004xk}, its conclusions for
situations with off-critical bulk were not correct.}.

The staircase connection naturally leads us to equations which
describe two-parameter
families of boundary perturbations, special cases of which
match the one-parameter flows found perturbatively in 
\cite{Fredenhagen:2009tn}, and which therefore also match the results
for fluctuating geometries found in \cite{Bourgine:2009zt}.
We expect that our more-general sets of flows from superpositions of 
boundary states can be generalised yet further, to describe
integrable bulk and boundary deformations of boundary 
conformal field theories with arbitrary numbers of boundary parameters. 
At the level of the exact $g$-function
equations the generalisation is rather clear, and will be indicated
below. However we will leave the detailed investigation of this point
for further work, as the two-parameter situation is already quite
involved. 

\resection{The staircase model, in the bulk and at the boundary}
\label{stairmodel}
The staircase model was originally introduced by Al.B.\,Zamolodchikov
in \cite{Zamolodchikov:1991pc}; various generalisations can be found
in \cite{Martins:1992ht,Dorey:1992bq,Martins:1992sx,Dorey:1992pj}.
Its S-matrix encodes the diagonal scattering of a single massive
particle of mass $M$, and can 
be obtained by the analytic continuation of the S-matrix of 
the sinh-Gordon model to those complex values of the coupling
constant where real-analyticity holds. At the level of the Lagrangian
the meaning of this continuation remains somewhat obscure, but as an
S-matrix theory the model appears to make perfect sense, and leads to
a consistent picture of finite-size effects described exactly by
thermodynamic Bethe ansatz (TBA)
equations which are derived in the standard way from the S-matrix.
Trading the analytically-continued sinh-Gordon coupling for a real
parameter $\tz$, the staircase S-matrix is
\beq
S(\theta)=
\tanh\left(\frac{\theta{-}\tz}{2}-\frac{i\pi}{4}\right)
\tanh\left(\frac{\theta{+}\tz}{2}-\frac{i\pi}{4}\right)
\label{stairS}
\eeq
and this leads to the following TBA equations for the pseudoenergy 
$\ep(\theta)$ for the system on a circle of circumference $R$\,:
\beq
\ep(\theta)=r\cosh\theta-\int_{\RR}\phi_S(\theta-\theta')L(\theta')d\theta'
\label{stairTBA}
\eeq
where $r=M\!R$, and
\beq
L(\theta)=\ln(1+e^{-\ep(\theta)})\,,\quad
\phi_S(\theta)=-\frac{i}{2\pi}\frac{d}{d\theta}\ln S(\theta)\,.
\eeq
The ground state energy of the system is then given by
$E(R)=-\frac{\pi}{6R}\,c_{\rm eff}(r)$, where the effective central charge
$c_{\rm eff}(r)$ is
\beq
c_{\rm eff}(r)=\frac{6}{\pi^2}\int_{\RR}r\cosh(\theta)L(\theta)\,d\theta\,.
\eeq

For later use we define
\beq
(x)(\theta)=
\frac%
{\sinh\left(\fract{\theta}{2}+\fract{i\pi x}{2}\right)}%
{\sinh\left(\fract{\theta}{2}-\fract{i\pi x}{2}\right)}~,\qquad
\phi_{(x)}(\theta)=-\frac{i}{2\pi}\frac{d}{d\theta}\ln\,(x)(\theta)=
\frac{-\sin(\pi x)/(2\pi)\,}{\cosh(\theta)-\cos(\pi x)}~,
\eeq
and also
\beq
\phi(\theta)=-\phi_{(\frac{1}{2})}(\theta)=\frac{1}{2\pi\cosh(\theta)}
\label{phidef}
\eeq
so that the staircase kernel $\phi_S(\theta)$ is
\beq
\phi_S(\theta)=\phi(\theta-\tz)+\phi(\theta+\tz)\,.
\label{phiS}
\eeq
For large values of $\tz$, this function is localised about
$\theta=\pm\tz$, as illustrated in figure \ref{phiSfig}.

\[
\begin{array}{c}
\includegraphics[width=0.95\linewidth]{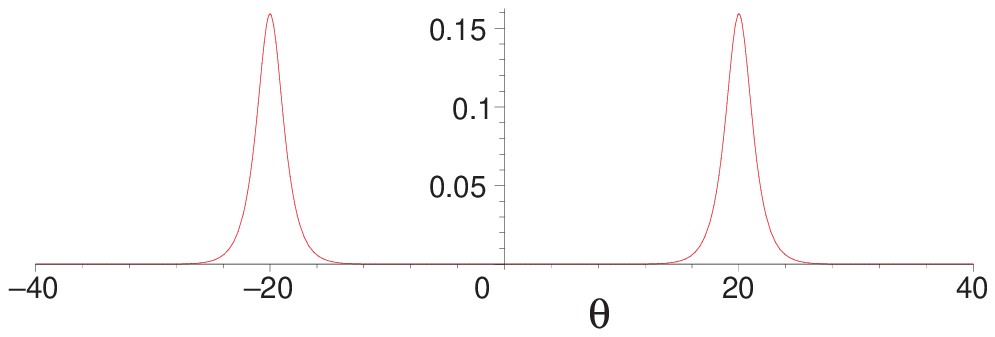}
\\
\parbox{0.7\linewidth}{
\small Figure \protect\ref{phiSfig}:
The staircase kernel $\phi_S(\theta)$ for $\tz=20$.
}
\end{array}
\]
\refstepcounter{figure}
\protect\label{phiSfig}

\noindent
The TBA system (\ref{stairTBA}) therefore couples the values of the
pseudoenergy $\ep(\theta)$ near to $\theta$ with those near to
$\theta\pm\tz$, and the behaviour of the effective central charge
$c_{\rm eff}(r)$ depends crucially on how many times the interval $[0,\tz]$
fits into the range $[{-}\ln(\frac{1}{r}),\ln(\frac{1}{r})]$\,, beyond
which the value of the pseudoenergy is dominated by the driving 
term $r\cosh\theta$ in (\ref{stairTBA}), irrespective of its coupling
to values taken elsewhere. Referring the reader to
\cite{Zamolodchikov:1991pc,Dorey:1992bq} for further explanation, the
net result is that $c_{\rm eff}(r)$ develops a series of plateaux, or steps,
which become more pronounced as $\tz\to\infty$. (The top curves
on figures \ref{gflows}a-\ref{gflows}d below show $c_{\rm eff}(r)$ for
$\tz=60$, by which point the plateaux are already quite sharply
defined.) Indexing the
steps by an integer $m=3$, $4$ \dots, the 
$(m{-}2)^{\rm th}$ step is found for $-(m{-}2)\tz/2\ll\ln(r)\ll
-(m{-}3)\tz/2$, and on this step, $c_{\rm eff}(r)\to c_m$ as
$\tz\to\infty$, where
\beq
c_m=1-\frac{6}{m(m{+}1)}
\eeq
is the central charge of the unitary minimal model $\CM_m$.
(More precisely, this holds in a double-scaling limit: pick 
$\bar{r}$ and $\bar\tz$ with
$-(m{-}2)\bar\tz/2<\ln(\bar r)< -(m{-}3)\bar\tz/2$, and
set $r=\bar r^{\rho}$ and $\tz=\rho\bar\tz$ with $\rho>0$;
then $\lim_{\rho\to\infty} c_{\rm eff}(r,\tz)=c_m$.) Furthermore, in the
crossover region $\ln(r)\approx -(m{-}3)\tz/2$ between the
$\CM_m$ and $\CM_{m-1}$ plateaux, 
the staircase pseudoenergy
$\ep(\theta)$ tends uniformly on suitably-shifted intervals
to the pseudoenergies $\ep_i(\theta)$ which solve 
the system of TBA equations
introduced in \cite{Zamolodchikov:1991vx} to describe the
$\phi_{13}$-induced flow from $\CM_{m}$ to $\CM_{m-1}$, a flow which
had previously been
found perturbatively in \cite{Zamolodchikov:1987ti,Ludwig:1987gs}.
(This interpolating theory was denoted by $\CM A^{(+)}_m$
in\cite{Zamolodchikov:1991vx}.)
This is illustrated in figure \ref{Lplots} for $m=4$.
\[
\begin{array}{c}
\includegraphics[width=0.95\linewidth]{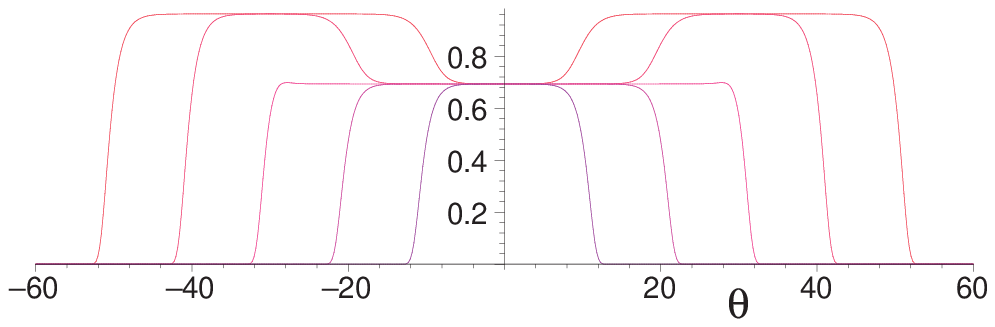}
\\[5pt]
\parbox{0.9\linewidth}{
\small Figure \protect\ref{Lplots}: The function
$L(\theta)$ at various values of $r$, for $\tz=60$.
From the upper to the lower curve, $\ln(r)$ is equal to $-50$,
$-40$, $-30$, $-20$ and $-10$. This range covers the crossover between
tricritical and critical Ising models, and the form of $L(\theta)$
around $\theta=+30$ and $-30$ approximates the functions
$\ln(1+e^{-\ep_1(\theta-30)})$ and
$\ln(1+e^{-\ep_2(\theta+30)})$ from
the corresponding interpolating TBA of
\cite{Zamolodchikov:1991vx} (cf.\ figure~1
of \cite{Dorey:2009vg}).
}
\end{array}
\]
\refstepcounter{figure}
\protect\label{Lplots}

The natural interpretation of these results -- also backed by
perturbative studies, in the spirit of \cite{Zamolodchikov:1987ti},
at large $m$ \cite{Lassig:1991ab} -- is that 
there is a one-parameter family of integrable quantum field theories 
with renormalisation group trajectories which, in the limit
$\tz\to\infty$, approach the union of the renormalisation
group trajectories of the $\CM A^{(+)}_m$ theories. From now on we
will assume this to be the case, and use it to deduce equations for
the flow of the $g$-function in these same theories. 

To treat the boundary staircase model using exact $g$-function
techniques, we need a conjecture for its boundary reflection factor
$R(\theta)$. It is natural to suppose that this can be obtained
through the same analytic continuation of the sinh-Gordon
boundary reflection factor as yielded the bulk S-matrix
(\ref{stairS}). The boundary sinh-Gordon model with no additional
boundary degrees of freedom has a two-parameter family of integrable
boundary conditions \cite{Ghoshal:1993tm}, and its reflection factor
follows from that of the first sine-Gordon breather, found in
\cite{Ghoshal:1993iq} (see also, for example, \cite{Corrigan:1999fp}).
Further continuing to the staircase values of the coupling, this leads
to the following reflection factor:
\beq
R(\theta)=\frac%
{(\frac{1}{2})(\frac{3}{4}-\frac{i\tz}{2\pi})
(\frac{3}{4}+\frac{i\tz}{2\pi})}%
{(\frac{1}{2}-\frac{E}{2})(\frac{1}{2}+\frac{E}{2})%
(\frac{1}{2}-\frac{F}{2})(\frac{1}{2}+\frac{F}{2})}
\label{R}
\eeq
where $E$ and $F$ are two parameters
whose relationship with the original two parameters of the boundary
sinh-Gordon model will not be relevant below. In the sinh-Gordon
model $E$ and $F$ are often real, but for the staircase model 
it will be more interesting to consider them at the complex values for
which real-analyticity is preserved, as with the 
continuation of the bulk coupling. Hence we set
\beq
E=\frac{i\tbi}{\pi}~,\qquad 
F=\frac{i\tbii}{\pi}
\eeq
with $\tbi$ and $\tbii$ real and, without loss of generality,
non-negative. There is an
obvious extension of this ansatz to incorporate $n$
boundary parameters $\theta_{b1}$ \dots $\theta_{bn}$\,:
\beq
R(\theta)=\frac%
{(\frac{1}{2})(\frac{3}{4}-\frac{i\tz}{2\pi})
(\frac{3}{4}+\frac{i\tz}{2\pi})}%
{\prod_{k=1}^{n}
(\frac{1}{2}-\frac{i\theta_{bk}}{2\pi})(\frac{1}{2}+\frac{i\theta_{bk}}{2\pi})}
\,.
\label{Rn}
\eeq
For $n>2$ this does not correspond to an integrable sinh-Gordon
boundary condition of the simple form treated in
\cite{Ghoshal:1993tm}, but it can be realised by the addition of a stack of
$n{-}2$ defects next to such a boundary \cite{Bowcock:2005vs,Bajnok:2007jg}.

For all of its subtleties at intermediate scales, the boundary
staircase model in the far infrared is simply a massive diagonal 
scattering theory, both in the bulk and at the boundary.
Its exact $g$-function should therefore be given by the 
formula proposed in \cite{Dorey:2004xk}. Explicitly, for a boundary
at the end of a cylinder of circumference $r=M\!R$,
\beq 
\ln g(r)= \ln\go(r)+\ln\gb(r)
\label{lng}
\eeq
where
\beq 
\ln\go(r)=
\sum_{n=1}^{\infty}
\frac{1}{2n}\int_{\RR^n}{\frac{d\theta_1}{1+e^{\epsilon(\theta_1)}}
\cdots\frac{d\theta_n}{1+e^{\epsilon(\theta_n)}}
\phi_S(\theta_1+\theta_2)\phi_S(\theta_2-\theta_3)\cdots
\phi_S(\theta_n-\theta_1)}
\label{lng0}
\eeq
and 
\beq 
\ln\gb(r)=\frac{1}{2}\int_{\RR}{d\te\left(\phi_b(\theta)-\phi_S(2\theta)-
\half\,\delta(\theta) \right) L(\theta)}
\eeq
where $\ep(\theta)$ solves the TBA equation (\ref{stairTBA}),
$L(\theta)=\ln(1+e^{-\ep(\theta)})$,
$\phi_S(\te)=-\frac{i}{2\pi}\frac{d}{d\theta}\ln
S(\theta)=\phi(\theta-\tz)+\phi(\theta+\tz)$, and,
restricting attention
to the $n=2$ case of (\ref{Rn}) for simplicity,
\begin{eqnarray} 
\phi_b(\te)
&=&-\frac{i}{2\pi}\frac{d}{d\theta}\ln R(\theta)\nn\\[4pt]
&=&
-\phi(\te)+
\phi_{(\frac{3}{4})}(\te-\half\tz)+
\phi_{(\frac{3}{4})}(\te+\half\tz)\nn\\[3pt]
&& {~~~~~}+\phi(\te-\tbi)+
\phi(\te+\tbi)+
\phi(\te-\tbii)+
\phi(\te+\tbii)\,.~~~
\end{eqnarray}
As in \cite{Dorey:2009vg}, this normalisation for $\phi_b$
differs by a factor of two from that used in
\cite{Dorey:2005ak,Dorey:2004xk}; we also used
$\phi(\theta)=-\phi_{(\frac{1}{2})}(\theta)$.
These equations are straightforward to implement numerically, and can
also be treated analytically in various limits. In the next section we
report some of the results of this analysis for the full staircase
model, while in section \ref{gMAplus} we show how the staircase
$g$-function equations decompose in suitable limits into sets of 
equations which govern exact $g$-function flows in the
interpolating theories $\CM A^{(+)}_m$.

\resection{The boundary staircase flows}
\label{sec3}
Following the split (\ref{lng}) of $\ln g$ into the sum 
$\ln\go+\ln\gb$, the single-integral piece
$\ln\gb$ naturally splits into three further terms as 
$\ln\gb= \ln\gi+ \ln\gii+ \ln\giii$\,,
with
\begin{align}
\ln\gi
&=
\frac{1}{2}\int_{\RR}{d\te\left(-\phi(\te)-
\half\delta(\theta)\right)\!L(\te)}\,,
\label{lngb1}\\
\ln\gii
&=
\frac{1}{2}\int_{\RR}{d\te\left(
\phi_{(\frac{3}{4})}(\te-\half\tz)+
\phi_{(\frac{3}{4})}(\te+\half\tz) - 
\phi(2\theta-\tz) -
\phi(2\theta+\tz)
\right)\!L(\te)}\,,
\label{lngb2}\\
\ln\giii
&=
\frac{1}{2}\int_{\RR}{d\te\left(\phi(\te-\tbi)+
\phi(\te+\tbi)+
\phi(\te-\tbii)+
\phi(\te+\tbii)\right)\!L(\te)}\,.
\label{lngb3}
\end{align}
Notice that $\ln\gi$ and $\ln\gii$ are independent of the boundary
parameters, and $\ln\gi$ only depends on the bulk parameter
$\tz$ implicitly, via the function $L(\theta)$.

In the large-$\tz$ limit, the full $g$-function 
passes through a series of plateaux as $\ln r$ varies, its value on each
plateau always matching  a (conformal) $g$-function value, or a
product of
such values, for the conformal field theory seen by the bulk theory
at that value of $r$. Some features of this behaviour can be seen
in figure~\ref{gflows}, where
for simplicity the values of $\tbi$ were chosen such that 
their associated boundary transitions always coincide with bulk
transitions. These and other aspects will be analysed in more detail
later.

\[
\begin{array}{c}
\\[-60pt]
\!\!
\!\!\!\!\!\!\!\!
\!\!\!\!\!\!\!\!
\includegraphics[width=0.64\linewidth]{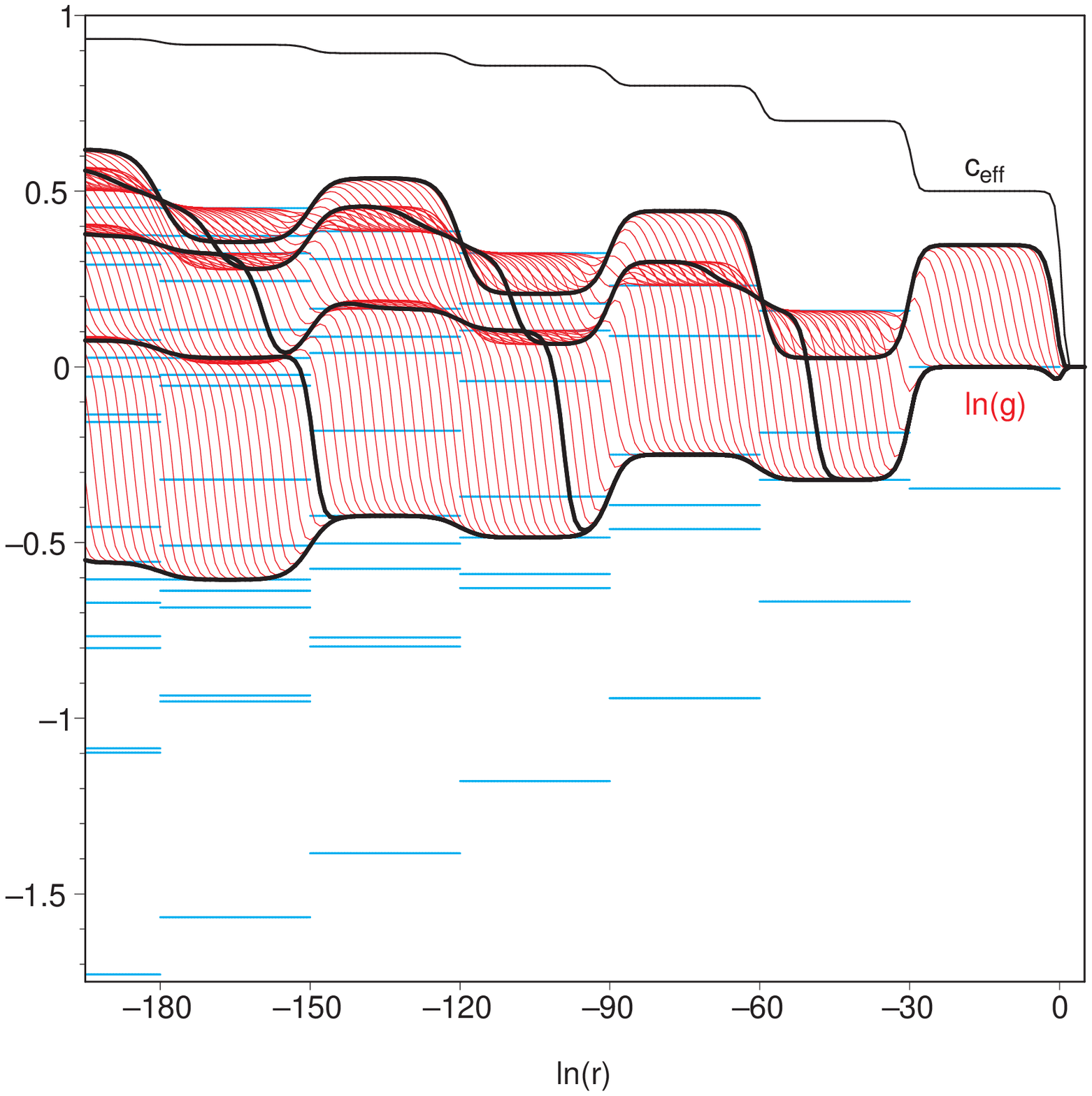}
\!\!\!\!
\!\!\!\!
\!\!\!\!\!\!\!\!
\includegraphics[width=0.64\linewidth]{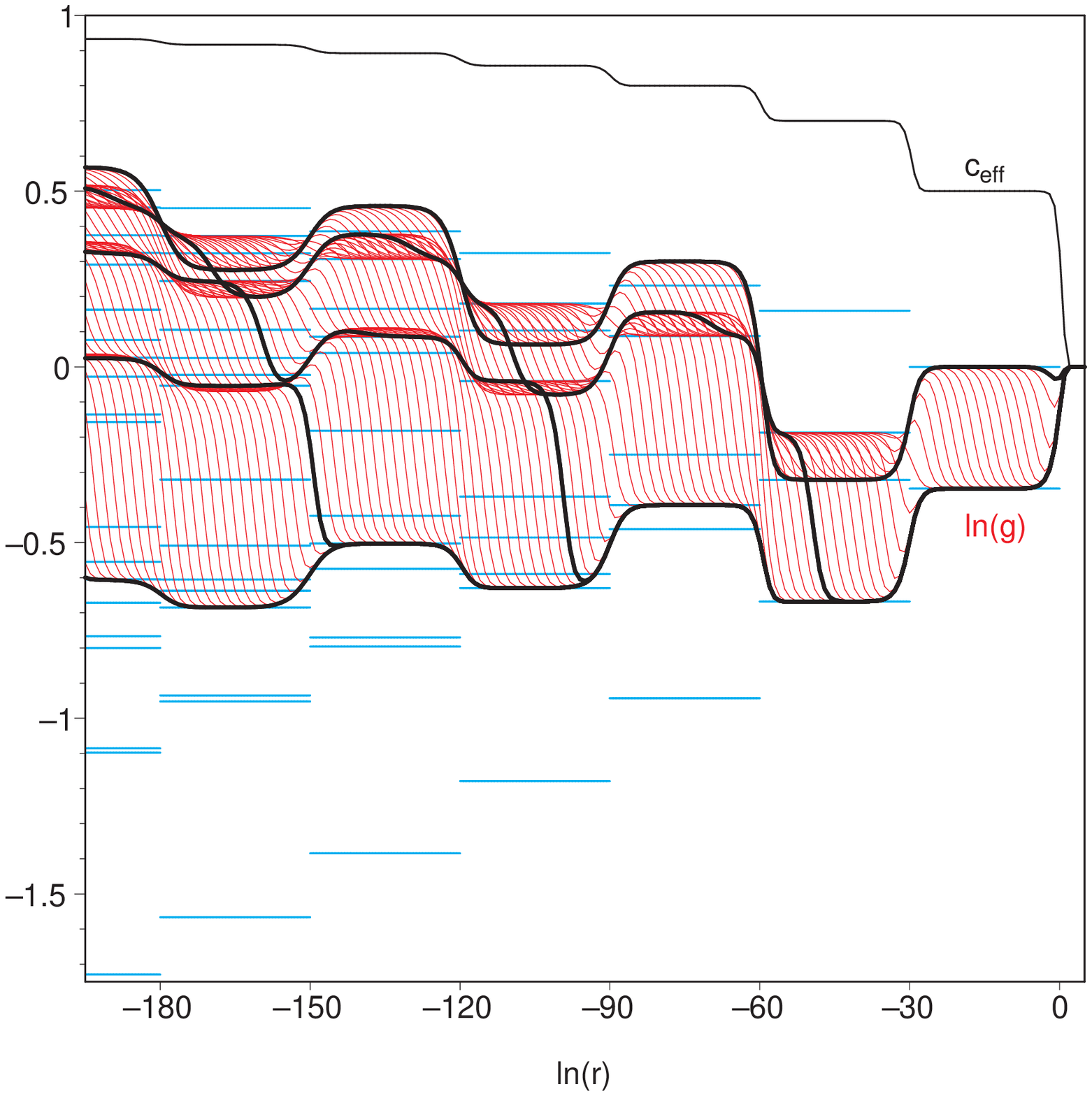}
\\
\!\!\!
\parbox{\linewidth}{\small Figure \ref{gflows}a: $\tbi=0$.\hskip
0.357\linewidth
Figure \ref{gflows}b: $\tbi=60$.}
\\
\!\!
\!\!\!\!\!\!\!\!
\!\!\!\!\!\!\!\!
\includegraphics[width=0.62\linewidth]{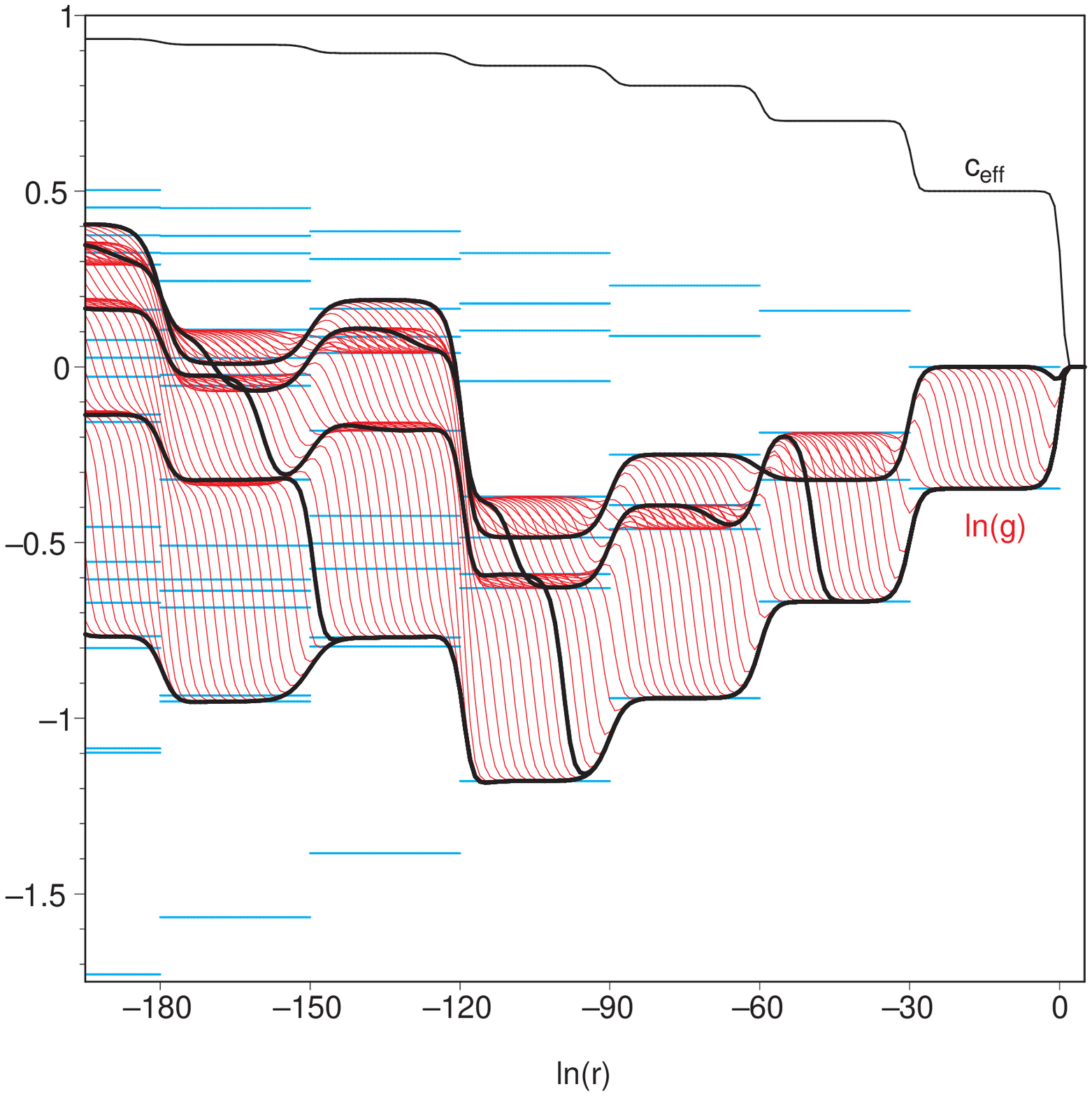}
\!\!\!\!
\!\!\!\!
\!\!\!\!\!\!\!\!
\includegraphics[width=0.62\linewidth]{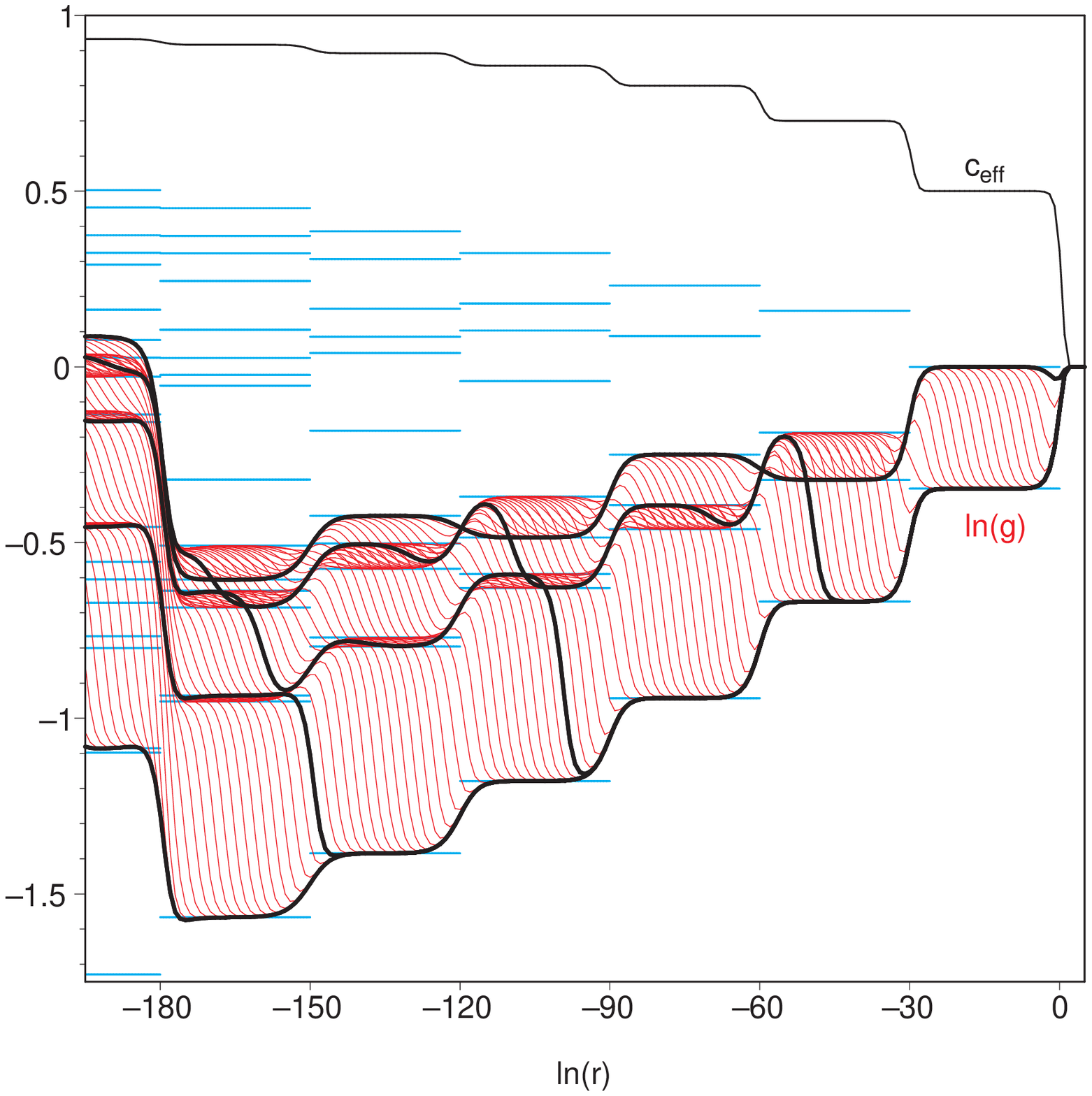}
\\
\!\!\!
\parbox{\linewidth}{\small Figure \ref{gflows}c: $\tbi=120$.\hskip
0.335\linewidth
Figure \ref{gflows}d: $\tbi=180$.}
\\[20pt]
\!\!\!\!\!
\!\!\!\!\!
\!\!\!\!\!\!
\parbox{0.98\linewidth}{
\small Figure \protect\ref{gflows}:
Staircase $g$-function flows for various values of $\tbi$. In
each plot $\tz=60$, and
the value of $\tbii$ is scanned from $0$ to $200$ in
steps of $2$. To aid the eye five $g$-function flows have been 
highlighted. The lowest-lying of them has $\tbii=200$; reading 
from left to right, this is then
joined by the flows for $\tbii=150$, 
$100$, $50$ and $0$ respectively. The short horizontal lines (light
blue online) show the logarithms of the conformal $g$-function values
(\ref{gmab})
for the `pure' (Cardy) boundaries of
each minimal model visited by the staircase flow. The other 
$g$-function plateaux
correspond to superpositions of these boundaries, as discussed in
the main text. 
}
\end{array}
\]
\refstepcounter{figure}
\protect\label{gflows}

\goodbreak

To give some precise formulae, we start by recalling some facts about
the minimal model $\CM_m$ and its conformal boundary conditions. 
The model has central charge
$c_m=1-\frac{6}{m(m{+}1)}$, and the permitted irreducible highest-weight
representations of the Virasoro algebra are labelled by pairs of
integers $a$ and $b$ with $1\le a\le m{-}1$ and $1\le b\le m$ subject
to the identifications $(a,b)\sim (m{-}a,m{+}1{-}b)$. They have
highest weight
\beq
h_{ab}=\frac{((m{+}1)a-mb)^2-1}{4m(m{+}1)}\,.
\label{hwr}
\eeq
In particular the bulk field $\phi_{13}$\,, which induces the
interpolating flow from $\CM_m$ to $\CM_{m-1}$\,, has scaling dimension 
$2h_{13}=2(m{-}1)/(m{+}1)$. The most general conformal boundary
condition is a superposition of `pure' (Cardy) boundary
conditions \cite{Cardy:1989ir}. There is one such boundary condition
for each irreducible highest-weight representation (\ref{hwr}), and
its boundary entropy (or $g$-function) is \cite{Affleck:1991tk}
\beq
g(m,a,b)=
\left(\frac{8}{m(m+1)}\right)^{\frac{1}{4}}\frac{\sin(\frac{a\pi}{m})
\sin(\frac{b\pi}{m+1})}{\sqrt{\sin(\frac{\pi}{m})\sin(\frac{\pi}{m+1})}}\,.
\label{gmab}
\eeq
Notice that this formula respects the identification $(a,b)\cong
(m{-}a,m{+}1{-}b)$, and has the additional symmetries
$g(m,a,b)=g(m,m{-}a,b)$ and $g(m,a,b)=g(m,a,m{+}1{-}b)$. 
This is the $\ZZ_2$ `spin flip' symmetry
of minimal model boundary conditions which 
in the case of the Ising model
 maps between spin up and spin down Dirichlet
boundaries.
For superpositions of Cardy boundaries, the
boundary entropies simply add.

We will also need some more detailed information about the form of the
function $L(\theta)$, illustrated for some sample values of $\ln(r)$
in figure \ref{Lplots}. Suppose
the bulk theory is near to the minimal model $\CM_m$, so that
$\ln(r)$ satisfies 
\beq
-(m{-}2)\tz/2 \ll \ln(r)\ll -(m{-}3)\tz/2 \,.
\label{rrange}
\eeq
Setting
$\alpha=2\ln(1/r)-(m{-}3)\tz$, we have
$0\ll\alpha\ll\tz$ and, starting from $\theta\approx \ln(1/r)$, 
$L(\theta)$ exhibits an alternating series
of plateaux of lengths $\alpha$ and $\tz-\alpha$, the
$i^{\rm th}$ plateau being centred at $\theta=z_i$, where
\beq
z_i=(m{-}2{-}i)\tz/2\,,\qquad i=1, 2,\dots 2m{-}5\,.
\eeq
(For later comparison with the TBA systems for $\CM A^{(+)}_m$ it will
be convenient to count these plateaux starting from the right.)
The plateau values of
$L(\theta)$ can be found as explained in 
\cite{Dorey:1992bq}. Adapting slightly the
notation from \cite{Zamolodchikov:1991vh,Zamolodchikov:1991vx}, 
define
constants $x_a$ and $y_a$ by
\beq
1+x_a=
\frac{\sin^2\Bigl(\frac{\pi a}{m+1}\Bigr)}%
{\sin^2\Bigl(\frac{\pi}{m+1}\Bigr)}~~,\qquad
1+y_a=
\frac{\sin^2\Bigl(\frac{\pi a}{m}\Bigr)}%
{\sin^2\Bigl(\frac{\pi}{m}\Bigr)}~.\qquad
\label{xy}
\eeq
Starting from $\theta=+\infty$, $L(\theta)$ is close to 
$0$ until $\theta\approx \ln(1/r)$, and then has 
height $\ln(1+x_2)$ for
$\ln(1/r)-\alpha\ll\theta\ll\ln(1/r)$, 
then $\ln(1+y_2)$ for $\ln(1/r)-\tz\ll\theta\ll\ln(1/r)-\alpha$, 
then $\ln(1+x_3)$ for
$\ln(1/r)-\tz-\alpha\ll\theta\ll\ln(1/r)-\tz$, 
and so on, before returning to $0$ for $\theta\ll-\ln(1/r)$. 
In full, the plateau values of $L(\theta)$ are
\begin{align}
\ln(1{+}x_{a}):&~~~~~~~~~~~
z_{2a-3}-\alpha/2\ll\theta\ll 
z_{2a-3}+\alpha/2\,,\qquad~~~~\, a=2\dots m{-}1\,;
\label{platx}\\[4pt]
\ln(1{+}y_{a}):&~~~\,
z_{2a-2}-(\tz{-}\alpha)/2\ll\theta\ll 
z_{2a-2}+(\tz{-}\alpha)/2\,,~~~\, a=2\dots m{-}2\,,
\label{platy}
\end{align}
and the complete sequence between $\theta=-\ln(1/r)$ and
$\theta=+\ln(1/r)$ is
\beq
\{
\,
\ln(1{+}x_{m-1}),\,
\ln(1{+}y_{m-2}),\,
\ln(1{+}x_{m-2})\,,
\ln(1{+}y_{m-3})\, \dots\,
\ln(1{+}y_{2}),\,
\ln(1{+}x_{2})
\,
\}\,.
\label{Ls}
\eeq
As a shorthand we will refer to the intervals in the set 
(\ref{platx}) as $x$-type, and those in the set (\ref{platy}) as
$y$-type.
From (\ref{xy}),
$\ln(1{+}x_1)= \ln(1{+}y_1)= 0$, and so
we can formally add these two constants to
the end of the sequence (\ref{Ls})
while remaining consistent with the values taken by 
$L(\theta)$ in the corresponding intervals, and likewise add 
$\ln(1{+}y_{m-1})$ and then $\ln(1{+}x_m)$ to the beginning. 
The symmetries $x_a=x_{m+1-a}$, $y_a=y_{m-a}$ reflect the more general
symmetry $L(\theta)=L(-\theta)$.

With these preliminaries completed we return to the exact $g$-function
$g(r)$. 
Three parts of $\ln g(r)$ do not
depend on the boundary parameters:
$\ln\go$, $\ln\gi$ and $\ln\gii$\,. These functions only undergo
transitions at the values of $\ln(r)$ where there is a bulk crossover,
that is at $\ln(r)=-(m{-}3)\tz/2$, $m=3$, $4$ \dots\,. The
effective equations governing these transitions in the
large-$\tz$ limit will be treated in the next section; here
instead we will suppose that
$r$ satisfies (\ref{rrange})
so that the bulk theory is close to the minimal model $\CM_m$.
Then $\go$,
$\gi$ and $\gii$ are approximately constant, and given by the
following formulae: 
%
%
\begin{numcases}{\ln g_0(r) = }
\displaystyle
{~}\ln\left(\left(\frac{8}{m(m+1)}\right)^{\frac{1}{4}}
\frac{\sin\frac{(m-1)\pi}{2m}}{\sqrt{\sin\frac{\pi}{m}\sin\frac{\pi}{m+1}}}
\right)&\mbox{for $m$ odd}
\label{lng0fodd}\\[10pt]
\displaystyle
{~}\ln\left(\left(\frac{8}{m(m+1)}\right)^{\frac{1}{4}}
\frac{\sin\frac{m\pi}{2(m+1)}}{\sqrt{\sin\frac{\pi}{m}\sin\frac{\pi}{m+1}}}
\right)&\mbox{for $m$ even}
\label{lng0feven}
\end{numcases} 

%
%
\begin{numcases}{\hskip -41pt\ln\gi(r)=-\half L(0)=}
-\half\ln(1{+}x_{(m+1)/2})=
-\half\ln\left(\frac{1}{\sin^2\frac{\pi}{m+1}}\right)
&\mbox{for $m$ odd}
\label{lngb1odd}
\\[10pt]
-\half\ln(1{+}y_{m/2})=
-\half\ln\left(\frac{1}{\sin^2\frac{\pi}{m}}\right)
&\mbox{for $m$ even}
\label{lngb1even}
\end{numcases} 

%
%
\begin{numcases}{\hskip -28pt\ln\gii(r)= -\half L\left(\half\tz\right)=}
-\half\ln(1{+}y_{(m-1)/2}) =
-\half
\ln\left(\frac{\sin^2\frac{(m-1)\pi}{2m}}{\sin^2\frac{\pi}{m}}\right) 
&\mbox{for $m$ odd}
\label{lngb2odd}
\\[10pt]
-\half\ln(1{+}x_{m/2}) =
-\half 
\ln\left(\frac{\sin^2\frac{m\pi}{2(m+1)}}{\sin^2\frac{\pi}{m+1}}\right)
&\mbox{for $m$ even}\,.
\label{lngb2even}
\end{numcases} 
These results are exact in the limit
$\{\,\tz\to\infty, 
-(m{-}2)\tz/2 \ll \ln(r)\ll -(m{-}3)\tz/2\, \}$.
The formula for $\ln\go$ will be derived in section~\ref{gMAplus} below.
Those for $\ln\gi$ and $\ln\gii$ follow from the fact that
$\phi(\theta)$ and $\phi_{(x)}(\theta)$ are only significantly
non-zero near to $\theta=0$. This means that the integrals (\ref{lngb1})
and (\ref{lngb2}) only 
receive contributions from the regions $\theta\approx 0$ and
$\theta\approx \tz/2$, where $L(\theta)$
is, by (\ref{platx}) and (\ref{platy}), 
approximately constant. Pulling $L(\theta)$ out
of each integral and using
\beq
\int_{\RR}{d\te\,\phi(\te)}=
-\int_{\RR}{d\te\,\phi_{(\frac{1}{2})}(\te)}=\frac{1}{2}\,,\qquad
\int_{\RR}{d\te\,\phi_{(\frac{3}{4})}(\te)}=-\frac{1}{4}
\eeq
together with the plateau values given by (\ref{platx}) and
(\ref{platy}) leads to (\ref{lngb1odd}) -- (\ref{lngb2even}).

When $\ln\go$, $\ln\gi$ and $\ln\gii$
are summed, the pieces which depend on whether
$m$ is odd or even cancel, leaving the following simple result, valid 
for all values of $m$:
\beq  
\ln\go+\ln\gi+\ln\gii=
\ln\left(\left(\frac{8}{m(m+1)}\right)^{\frac{1}{4}}
\sqrt{\sin\frac{\pi}{m}\sin\frac{\pi}{m+1}}\right)\,.
\label{lng0sum}
\eeq
This is the logarithm of $g(m,1,1)$ or $g(m,m{-}1,1)$, the boundary 
entropy of the
conformal boundary condition associated with the bulk vacuum field or
its $\ZZ_2$ spin flip conjugate.
In fact it is not surprising that this partial sum should be equal to 
the logarithm of
the full boundary entropy for some boundary condition, as will become
clear as we examine the behaviour of the remaining part of $\ln g$,
namely $\ln\giii$\,.

Since $\ln\giii$ depends on the boundary parameters, we would expect it
to undergo transitions not just where the bulk
crossovers occur, but also, possibly, at energy scales related to
pure-boundary transitions, and this turns out to be the case.
The integral (\ref{lngb3}) receives
contributions from 
$\theta\approx\pm\tbi$ and
$\theta\approx\pm\tbii$\,; if these regions lie within the
$x$- and $y$-type
intervals (\ref{platx}) and (\ref{platy}) 
then, pulling
$L(\theta)$ outside the integrals and recalling that
$L(\theta)=L(-\theta)$,
\beq
\ln\giii(r)=
\half\left(L\left(\tbi\right)+L\left(\tbii\right)\right)
 \label{lngb3f}
\eeq
and the value of $\ln\giii(r)$ will not change for small changes in
$r$. 
Conversely, $\ln\giii(r)$ will undergo a crossover whenever $r$
is such that 
either $\tbi$ or $\tbii$ lies on a boundary between the
intervals (\ref{platx}) and (\ref{platy}).
This means that there
will be boundary transitions associated with the parameter
$\tbi$ at
\beq
\ln(r)=-k\tz-\tbi\,,\quad  k=0,1,\dots
\eeq
and 
\beq
\ln(r)=-k\tz+\tbi\,,\quad  k=A,A{+}1,\dots
\eeq
where $A=\lceil 2\tbi/\tz\rceil$, the smallest integer
greater than or equal to $2\tbi/\tz$\,. An analogous
formula holds for $\tbii$. 

If $\ln(1/r)$ is smaller than $\tbi$, $L(\theta)$ is
effectively zero near to $\theta=\pm\tbi$ and the term
$\frac{1}{2}L(\tbi)$ ceases to contribute to $\ln\giii(r)$.
This explains why the plots in figure \ref{gflows} stabilise with
increasing $\tbi$, with the parts of the plots
with $\ln(r)>-\tbi$ being independent of $\tbi$.
If $\ln(1/r)$ is smaller than $\tbi$ and $\tbii$,
then $\ln\giii(r)$ is zero and the logarithm of the
full $g$-function is given by the
sum $\ln\go(r)+\ln\gi(r)+\ln\gii(r)$, which we already observed was equal to
$\ln g(m,1,1)$ when the bulk theory is on the plateau
corresponding to $\CM_m$. 
If the limit $\tbi\to\infty$,
$\tbii\to\infty$ is taken {\em before}\/ $r$ is varied, we 
see that there is one flow which simply moves
through the $(1,1)$ boundary conditions in the successive minimal
models, its $g$-function being given by
$\ln g(r)=\ln\go(r)+\ln\gi(r)+\ln\gii(r)$ for {\em all}\/ values of $r$.
(This is why the partial sum
(\ref{lng0sum}) is itself the logarithm of a boundary entropy.)
For $\ln(r)>-180$, this flow is matched by the lowest-lying curve of figure 
\ref{gflows}d. Notice that since $\ln\giii(r)$ is manifestly positive,
all other $g$-function flows must lie above this limiting curve, an
off-critical generalisation of the fact that at a fixed point the
lowest-possible boundary entropy is always found for the $(1,1)$
boundary condition.

For smaller values of $\tbi$ and $\tbii$ the picture
becomes more complicated, as can already be seen from figure
\ref{gflows}. Nevertheless it is still possible to formulate general
rules for the boundary conditions which are visited.
The plateau behaviour of $L(\te)$ means that the sequence of
boundary conditions seen for any given $\tbi$ and $\tbii$ depends 
on the intervals 
(\ref{platx}) and (\ref{platy})
that they (and their negatives) find themselves in as $\ln(r)$ varies.
From (\ref{Ls}) and the immediately-following remarks, the
possible values of $L(\te)$ on these intervals
are
the elements of the set
\beq
 \left\{\ln(1+x_a), \ln(1+y_b)\right\} 
\label{lxy}
\eeq
where the indices $a$ and $b$ lie in the ranges
\beq
a\in\{1,\cdots,m\},\;b\in\{1,\cdots,m-1\}
\label{ab}
\eeq
and the value of $0$, found for $|\theta|\gg\ln(1/r)$,
arises when $a$ is equal to $1$ or $m$, or $b$ is equal to $1$ or
$m{-}1$.
The symmetries $x_{m+1-a}=x_a$, $y_{m-b}=y_b$ could have been used 
to restrict the indices $a$ and $b$ to 
\begin{eqnarray}
a\in\{1,\cdots,\frac{m+1}{2}\},\;b\in\{1,\cdots,\frac{m-1}{2}\}\quad
&\mbox{for m odd}&\nonumber\\
\mbox{and}\qquad\qquad\qquad\qquad\qquad{a,b}\in\{1,\cdots,\frac{m}{2}\}
\quad &\mbox{for m even}&
\label{abr}
\end{eqnarray}
but for reasons to be explained below it will be convenient to keep
with the larger ranges.
A given pair of boundary parameters corresponds to two (possibly
equal) values of $L(\theta)$ from the set (\ref{lxy}), and, 
via the exponential of
(\ref{lngb3f}) and (\ref{lng0sum}), to a value for the
$g$-function. We found that this value can always be expressed as a
sum of Cardy $g$-function values (\ref{gmab}), according to the
following rules, where we introduce the convenient notation
$[x_a,y_b]$ and so on to denote the particular combinations of boundary
conditions which arise: 
\beq
\!\!\!
\begin{array}{lll}
 L(\tbi)&L(\tbii)&\mbox{Boundary condition}\\[3pt]
\ln\left(1{+}x_a\right)&\ln\left(1{+}y_b\right)&
\rlap{$[x_a,y_b]$}\phantom{[x_p,x_q]}
\equiv
(b,a)\\
\ln\left(1{+}x_p\right)&\ln\left(1{+}x_q\right)&
[x_p,x_q]\equiv
(1,|p{-}q|{+}1)\&
(1,|p{-}q|{+}3)\&
\cdots\&
(1,m{-}|p{+}q{-}m{-}1|)\\
\ln\left(1{+}y_r\right)&\ln\left(1{+}y_s\right)& 
\rlap{$[y_r,y_s]$}\phantom{[x_p,x_q]}
\equiv
(|r{-}s|{+}1,1)\&
(|r{-}s|{+}3,1)\&
\cdots\&
(m{-}1{-}|r{+}s{-}m|,1)\\
\end{array}
\label{gident}
\eeq
with the same result for $\tbi\leftrightarrow\tbii$. Notice that if
all indices are restricted to the reduced ranges
(\ref{abr}), the rules simplify to
\beq
\begin{array}{lll}
 L(\tbi)&L(\tbii)&\mbox{Boundary condition}\\[3pt]
\ln\left(1+x_a\right)&\ln\left(1+y_b\right)&(b,a)\\
\ln\left(1+x_p\right)&\ln\left(1+x_q\right)&
(1,|p{-}q|{+}1)\&
(1,|p{-}q|{+}3)\&
\cdots\&
(1,p{+}q{-}1)\\
\ln\left(1+y_r\right)&\ln\left(1+y_s\right)& 
(|r{-}s|{+}1,1)\&
(|r{-}s|{+}3,1)\&
\cdots\&
(r{+}s{-}1,1)\,.
\end{array}
\label{gidentr}
\eeq
The identifications of $g$-function values with specific boundary
conditions implied by (\ref{gident}) should be treated with care for a
couple of reasons. First, ambiguity arises from the equalities
$g(m,a,b)=g(m,m{-}a,b)$, $g(m,a,b)=g(m,a,m{+}1{-}b)$.
These are related to the symmetries 
$L(\theta)=L(-\theta)$, $x_a=x_{m+1-a}$, $y_a=y_{m-a}$
of $L(\theta)$ and its plateau values, and, more precisely, to
the following symmetries of the boundary condition combinations 
given in (\ref{gident}):
\begin{align}
[x_a,y_b]&= [x_{m+1-a},y_{m-b}] &
\overline{[x_a,y_b]}&=
[x_{m+1-a},y_{b}]= [x_{a},y_{m-b}] \nn\\[3pt]
[x_p,x_q]&= [x_{m+1-p},x_{m+1-q}]&
\overline{[x_p,x_q]}&=
[x_{m+1-p},x_{q}]= [x_{p},x_{m+1-q}] \label{spinflip} \\[3pt]
[y_r,y_s]&= [y_{m-r},y_{m-s}]&
\overline{[y_r,y_s]}&=
[y_{m-r},y_{s}]= [y_{r},y_{m-s}] \nn
\end{align}
where the overbar denotes the $\ZZ_2$ `spin flip' symmetry
mentioned just after (\ref{gmab}), acting on individual Cardy
boundaries as $\overline{(a,b)}=(m{-}a,b)\equiv (a,m{+}1{-}b)$\,. 
As remarked in~\cite{Dorey:2009vg}, it should be possible 
to resolve such ambiguities in a systematic fashion
by studying the renormalisation group flows
of the inner products of finite-volume 
excited states~\cite{Bazhanov:1996aq,Dorey:1996re}
with the
boundary states, but we shall leave this for future work.
Second, there are sometimes further, more accidental,
degeneracies in the set of non-negative-integer sums of 
Cardy $g$-function values -- for example,
$g(5,1,3)=2g(5,1,1)$, so that in $\CM_5$ the $(1,3)$ and
$(1,1)\&(1,1)$ boundary conditions cannot be distinguished by their
$g$-function values alone. Nevertheless, and modulo the spin flip
ambiguity just described, (\ref{gident}) is the only set 
of decompositions we have found which works in a uniform fashion for all
$m$. From now on we shall assume that it is correct, and mostly leave
the spin flip ambiguity implicit.

\medskip

Continuing to suppose that the bulk theory is in 
the vicinity of the bulk fixed point $\CM_m$, we now
let $\ln(r)$ vary from the lower to the upper end of
the range (\ref{rrange}), that is from 
$-(m{-}2)\tz/2$ to $-(m{-}3)\tz/2$\,. 
The centres of the intervals (\ref{platx}) and (\ref{platy}) 
remain fixed, at $\theta=z_i$, $i=1\dots 2m{-}5$,
but the widths of the $x$-type intervals,
$\alpha\equiv 2\ln(1/r)-(m{-}3)\tz$,
decrease from $\tz$ to zero, while those of the $y$-type intervals,
$\tz-\alpha$, increase from zero to $\tz$. Thus so long as
$|\tbi|< (m{-}2)\tz/2$ and $\tbi$ 
is not an integer multiple of $\tz/2$, the regions
$\theta\approx\pm\tbi$ 
move from $x$-type intervals to $y$-type
intervals during this process, and the value of $L(\pm\tbi)$, and hence
that of the $g$-function, undergoes a change. For brevity we will
phrase the rest of the discussion in terms of the intervals seen by
$L(\tbi)$, but we could equally look at
$L(-\tbi)$. By the $\ZZ_2$ ambiguity just discussed this
might lead to different boundary conditions being assigned but since
the $g$-functions are blind to their difference, we will ignore
this issue for now. In fact, it will be convenient to allow $\tbi$
(and $\tbii$) to take positive and negative values, so all options are
in any case covered.

Suppose, then, that at the start of the process $\tbi$ is
in the $x$-type interval centred at
$\theta=z_{2r-3}$, and $\tbii$ is in the $x$-type interval centred
at $\theta=z_{2s-3}$, corresponding to plateau values for
$L(\tbi)$ and $L(\tbii)$
equal to $\ln(1+x_r)$ and $\ln(1+x_s)$ respectively.
The $y$-type interval
that $\tbi$ moves to depends on its position relative to $z_{2r-3}$.
If $\tbi>z_{2r-3}$ then 
$L(\tbi)$ moves to the plateau $\ln(1+y_{r-1})$, while if
$\tbi<z_{2r-3}$ it moves to the plateau $\ln(1+y_r)$.
An identical set of possibilities occurs for $L(\tbii)$, with a
transition which may occur before of after that in $L(\tbi)$
depending on the relative distances of $\tbi$ and $\tbii$ from the
centres of their original ($x$-type) intervals. Putting these
ingredients together gives the following `skeleton' of transitions 
from an initial situation where
$L(\tbi)=\ln(1{+}x_r)$ and $L(\tbii)=\ln(1{+}x_s)$\,:
\[
\begin{CD}
[y_{r},y_{s}] @<<< [x_r,y_{s}] @>>> [y_{r-1},y_{s}] \\[2pt]
@AAA @AAA @AAA\\[2pt]
[y_{r},x_{s}] @<<< [x_r,x_{s}] @>>> [y_{r-1},x_{s}] \\[2pt]
@VVV @VVV @VVV\\[2pt]
[y_{r},y_{s-1}] @<<< [x_r,y_{s-1}] @>>> [y_{r-1},y_{s-1}] 
\end{CD}
\]
This diagram encapsulates the boundary flows seen while the bulk is in
the vicinity of $\CM_m$, if $\tbi$ and $\tbii$ are in the intervals
$[z_{2r-3}-\tz/2,z_{2r-3}+\tz/2]$,
$[z_{2s-3}-\tz/2,z_{2s-3}+\tz/2]$. 
The complete set of
flows seen near to $\CM_m$ is the union of a number of such diagrams,
so as to cover the full ranges $|\tbi|\le (m{-}2)\tz/2$,
$|\tbii|\le (m{-}2)\tz/2$. To give one example, at $m=5$ this
results in a $7\times 7$ grid, the lower ($\tbii\ge 0$) half of which
is:
\[
\begin{array}{ccccccccccccc}
\vdots &&\vdots &&\vdots &&\vdots &&\vdots &&\vdots &&\vdots \\[2pt]
[y_4,x_3] &\larr& [x_4,x_3] &\rarr& 
[y_3,x_3] &\larr& [x_3,x_3] &\rarr& 
[y_2,x_3] &\larr& [x_2,x_3] &\rarr& 
[y_1,x_3] \\[10pt]
\Big\downarrow&& \Big\downarrow&& \Big\downarrow&& \Big\downarrow&&
\Big\downarrow&& \Big\downarrow&& \Big\downarrow\\[10pt]
[y_4,y_2] &\larr& [x_4,y_2] &\rarr& 
[y_3,y_2] &\larr& [x_3,y_2] &\rarr& 
[y_2,y_2] &\larr& [x_2,y_2] &\rarr& 
[y_1,y_2] \\[10pt]
\Big\uparrow&& \Big\uparrow&& \Big\uparrow&& \Big\uparrow&&
\Big\uparrow&& \Big\uparrow&& \Big\uparrow\\[10pt]
[y_4,x_2] &\larr& [x_4,x_2] &\rarr& 
[y_3,x_2] &\larr& [x_3,x_2] &\rarr& 
[y_2,x_2] &\larr& [x_2,x_2] &\rarr& 
[y_1,x_2] \\[10pt]
\Big\downarrow&& \Big\downarrow&& \Big\downarrow&& \Big\downarrow&&
\Big\downarrow&& \Big\downarrow&& \Big\downarrow\\[10pt]
[y_4,y_1] &\larr& [x_4,y_1] &\rarr& 
[y_3,y_1] &\larr& [x_3,y_1] &\rarr& 
[y_2,y_1] &\larr& [x_2,y_1] &\rarr& 
[y_1,y_1] 
\end{array}
\]

\bigskip

The entries in square brackets can be converted into specific
boundary conditions by 
using the dictionary (\ref{gident}), to give the picture shown in
figure \ref{m=5} below. We note once more that these predictions are
made modulo the $\ZZ_2$ ambiguity in the relationship between
$g$-function values and boundary conditions. The options chosen here,
which follow from the rule formulated earlier,
are consistent with predictions made in, for example, 
\cite{Graham:2001pp,Fredenhagen:2009tn}, but we have not attempted to
confirm them using exact $g$-function techniques. As already
mentioned, this would require the computation of inner products of
states other than the ground state with the boundary state. 

Notice
that the figure is symmetrical about the diagonals $\tbi=\tbii$
and $\tbi=-\tbii$, while negating either $\tbi$ or $\tbii$ individually
has the same effect as the $\ZZ_2$ spin flip.
This second feature means that the boundary conditions in the middle
column on the figure, $\tbi=0$ (or 
equivalently, the top row shown, $\tbii=0$) are mapped 
into themselves under the
spin flip.
As follows from (\ref{spinflip}), the corresponding
properties hold for all other values of
$m$, if it is assumed that the rules (\ref{gident}) are correct
in general.
\[
\begin{array}{c}
\hskip -20pt
\includegraphics[width=1.05\linewidth]{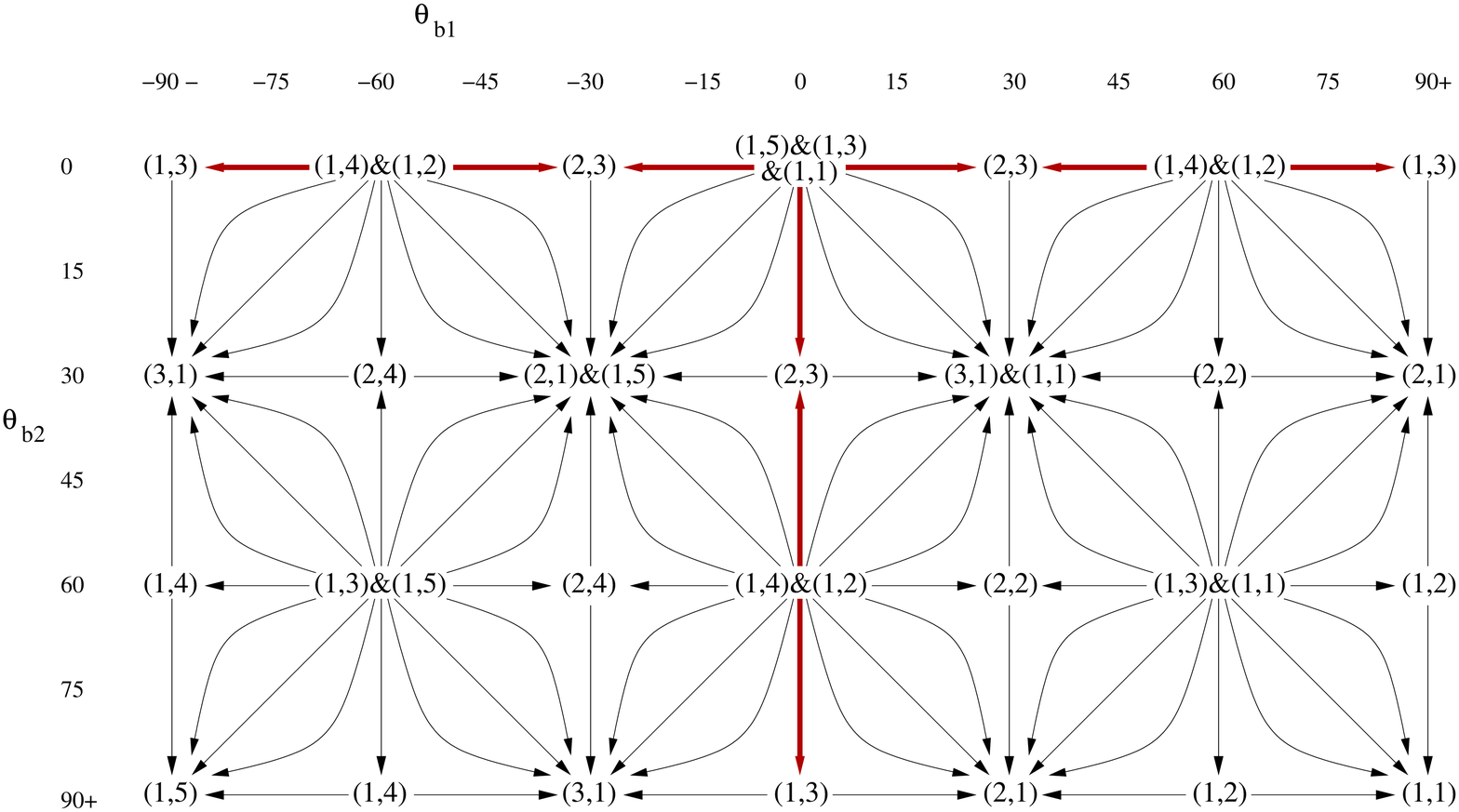}
\\[11pt]
\parbox{0.99\linewidth}{
\small Figure \protect\ref{m=5}: A diagram of the pure boundary
flows appearing near to the bulk fixed point $\mathcal{M}_5$, 
for $\tz=60$ and various values of $\tbi$ and $\tbii$. The
slightly-thicker vertical and horizontal arrows (red online) indicate
the flows and boundary conditions which are mapped to themselves under
the $\ZZ_2$ spin flip.
}
\end{array}
\]
\refstepcounter{figure}
\protect\label{m=5}

\medskip

To complete the picture we must examine what happens when the bulk
theory moves between two neighbouring fixed points, say $\CM_m$ and
$\CM_{m-1}$. The bulk
flow occurs when $\ln(r)$ varies through a region
centred on $\ln(r)= -(m{-}3)\tz/2$ which is 
of size of order one as $\tz\to\infty$.
As this
happens the form of $L(\theta)$ changes: the $x$-type plateaux for
$\CM_m$ shrink to zero size, while the $y$-type plateau of
height $\ln(1+y_a)|_{\CM_m}$, $a=1\dots m{-}1$, for $\CM_m$
becomes the $x$-type plateau $\ln(1+x_a)|_{\CM_{m-1}}$
for $\CM_{m-1}$, with $y$-type plateaux for $\CM_{m-1}$
opening up between these as $\ln(r)$ increases further.
(By (\ref{xy}), $\ln(1+y_a)|_{\CM_m}=\ln(1+x_a)|_{\CM_{m-1}}$, so 
there is no sudden change in heights implied by the
redesignation of $y$-type to $x$-type plateaux over the transition.)

Now consider the behaviour of the logarithm of the full
$g$-function, $\ln g=\ln\go+\ln\gi+\ln\gii+\ln\giii$, as $\CM_m$ flows
to $\CM_{m-1}$. From 
(\ref{lng0sum}), the sum of the first three terms,
$\ln\go+\ln\gi+\ln\gii$, changes from $\ln g(m,1,1)$ to $\ln
g(m{-}1,1,1)$. The behaviour of the remaining piece, $\ln\giii$,
depends on the values of $\tbi$ and $\tbii$. Suppose first that
both $|\tbi|\le (m{-}3)\tz/2$ and 
$|\tbii|\le (m{-}3)\tz/2$, so that neither boundary parameter has 
become decoupled at the point of the bulk transition under discussion. 
There are then three cases:

\smallskip

\noindent {\bf (i)}
If neither $\tbi$ nor $\tbii$ are at the centre of
what had been an $x$-type plateau for $\CM_m$, that is if
\beq
\tbi, \tbii \neq (m{-}3)\tz/2-k\tz\,,\qquad k=0,1,\dots m-3
\eeq
then both $\tbi$ and $\tbii$ will have undergone a
`pure-boundary' transition of the sort described earlier, from an
$x$-type onto a $y$-type plateau,
{\em before}\/ the $\CM_m\to\CM_{m-1}$ transition is
reached. Thus the conformal boundary condition seen just before
the bulk transition occurs will correspond to some pair
$[y_r,y_s]|_{\CM_m}$ where $2\le r,s\le m{-}2$, corresponding to
$\tbi$ and $\tbii$ lying in the intervals
$((m{-}3)\tz/2{-}(r{-}1)\tz\,,\,(m{-}3)\tz/2{-}(r{-}2)\tz)$ and
$((m{-}3)\tz/2{-}(s{-}1)\tz\,,\,(m{-}3)\tz/2{-}(s{-}2)\tz)$\,.
After the transition the plateau values will
not have changed but their interpretations will have, to the pair
$[x_r,x_s]|_{\CM_{m-1}}$\,. Translated into specific conformal
boundary conditions using (\ref{gident}) at $m$ and $m-1$ the flow
is therefore
\beq
\begin{array}{cl}
(f,1)\,\&\,
(f{+}2,1)\,\&\,
\cdots\,\&\,
(g,1) & ~~\CM_m\\[4pt]
\downarrow & \\[5pt]
(1,f)\,\&\,
(1,f{+}2)\,\&\,
\cdots\,\&\,
(1,g) & ~~\CM_{m-1}
\end{array}
\label{case1}
\eeq
where
\beq
f=|r{-}s|{+}1~,\quad g=m{-}1{-}|r{+}s{-}m|
\quad \mbox{and}~~ 2\le r,s \le m{-}2\,.
\label{rsrange}
\eeq
Via (\ref{spinflip}) and the symmetry under $\tbi\leftrightarrow\tbii$ 
the full set of options is explored by restricting $r$ and $s$ to the
fundamental domain $2\le r\le s\le m-2$, $r+s\le m$. In fact,
(\ref{rsrange}) is equivalent to $f$ and $g$ in (\ref{case1})
being restricted by
\beq
1\le f<g\le m-1\,,\quad f-g\in 2\ZZ\,.
\eeq
Notice that these, the generic flows, always start at `sinks'
on networks of pure-boundary flows such as figure \ref{m=5}, and end
on `sources' on the corresponding network one minimal model down.

\smallskip

\noindent {\bf (ii)}
If one of $\tbi$ or $\tbii$ lies at the centre of an $x$-type plateau
for $\CM_m$, then it remains on that plateau right up to 
the moment of the bulk transition, after which it will instead lie in
the centre of a $y$-type plateau for $\CM_{m-1}$.
If this centre is located at $\theta=(m-3)\tz/2-(s{-}2)\tz$, 
$2\le s\le m{-}1$, then the corresponding
value of $L(\theta)$ moves from $\ln(1{+}x_s)|_{\CM_m}$ to
$\ln(1{+}y_{s-1})|_{\CM_{m-1}}$\,. 
The other plateau value simply changes its designation
from $\ln(1{+}y_r)|_{\CM_m}$ to
$\ln(1{+}x_r)|_{\CM_{m-1}}$\,, as in case~{\bf (i)}. 
Thus the boundary condition flow is
$[x_s,y_r]|_{\CM_m}\to [y_{s-1},x_r]_{\CM_{m-1}}$, or
\beq
\begin{array}{cl}
(r,s) & ~~\CM_m\\[4pt]
\downarrow & \\[5pt]
(s{-}1,r) & ~~\CM_{m-1}
\end{array}
\label{case2}
\eeq
where
\beq
2\le r\le m{-}2\,,\quad
2\le s\le m{-}1\,.
\eeq

\smallskip

\noindent {\bf (iii)} 
Lastly, if both $\tbi$ and $\tbii$ are at the centres of $x$-type
plateaux for $\CM_m$, say at
$(m-3)\tz/2-(r{-}2)\tz$ and
$(m-3)\tz/2-(s{-}2)\tz$ with $2\le r,s\le m{-}1$, 
then reasoning as above the boundary condition flow is
$[x_r,x_s]|_{\CM_m}\to [y_{r-1},y_{s-1}]|_{\CM_{m-1}}$, or
\beq
\begin{array}{cl}
(1,f)\,\&\,(1,f{+}2)\,\&\,\cdots\,\&\,(1,g) & ~~\CM_{m}\\[4pt]
\downarrow & \\[5pt]
(f,1)\,\&\,(f{+}2,1)\,\&\,\cdots\,\&\,(g{-}2,1) & ~~\CM_{m-1}
\end{array}
\eeq
where this time
\beq
f=|r{-}s|{+}1~,\quad g=m{-}|r{+}s{-}m{-}1|
\quad \mbox{and}~~ 2\le r,s \le m{-}1\,,
\eeq
or equivalently
\beq
1\le f<g\le m\,,\quad f-g\in 2\ZZ\,.
\eeq
For these (least-generic) cases the flows are always from sources to 
sinks on neighbouring pairs of pure-boundary
networks such as figure \ref{m=5}, and decrease by 
one the number of superposed Cardy states.

\medskip

Finally we must treat the cases where either one or both of $|\tbi|$ 
and $|\tbii|$ is larger than $(m{-}3)\tz/2$\,. Then the corresponding
plateau values of $L(\theta)$ simply flow from zero to zero.
If in addition 
neither $\tbi$ nor $\tbii$ lie at the centre of an $x$-type plateau,
then it is easily seen that the situation is covered by case 
{\bf (i)} above, if the indices $r$ and $s$ are allowed to
take the additional values of $1$ and $m{-}1$ (recall that
$\ln(1{+}y_1)|_{\CM_m}=\ln(1{+}y_{m-1})|_{\CM_m}=
\ln(1{+}x_1)|_{\CM_{m-1}}=\ln(1{+}x_{m-1})|_{\CM_{m-1}}=0$). Thus the
combined story is that there are flows of the form (\ref{case1}) for
every pair $(f,g)$ satisfying
\beq
1\le f\le g\le m-1\,,\quad f-g\in 2\ZZ\,.
\eeq
(Taking $f=g=1$ or $f=g=m{-}1$ gives the flows 
$(1,1)|_{\CM_m}\to (1,1)|_{\CM_{m-1}}$ and
$(m{-}1,1)|_{\CM_m}\to (1,m{-}1)|_{\CM_{m-1}}$ which occur when both
$|\tbi|$ and $|\tbii|$ are larger than $(m{-}3)\tz/2$, so that $\ln\giii$
remains zero throughout the flow.) Last of all, if, say,
$|\tbii|>(m{-}3)\tz/2$ while $\tbi$ is at
the centre of an $x$-type plateau, then the flow is
$[y_1,x_s]|_{\CM_m}\to [x_1,y_{s-1}]|_{\CM_{m-1}}$
or $[y_{m-1},x_s]|_{\CM_m}\to [x_{m-1},y_{s-1}]|_{\CM_{m-1}}$ with
$2\le s\le m{-}1$, which simply means that the indices in
(\ref{case2}) can be given the enlarged range
\beq
1\le r\le m{-}1\,,\quad
2\le s\le m{-}1\,.
\eeq

\[
\begin{array}{c}
\includegraphics[width=0.85\linewidth]{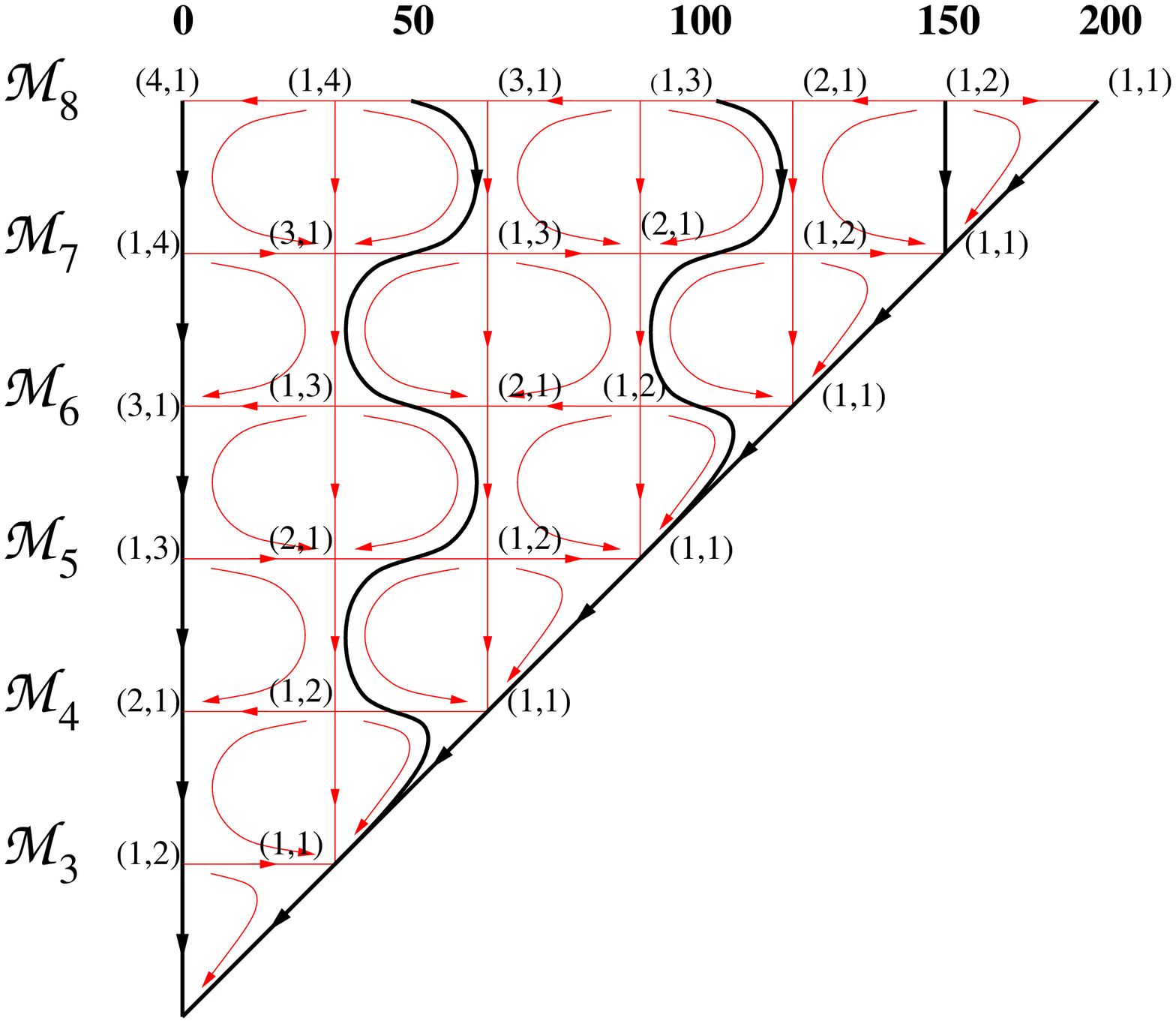}
\\[5pt]
\parbox{0.95\linewidth}{
\small Figure \protect\ref{firstflows}: A depiction of the flows 
corresponding
to figure \ref{gflows}d, for which $\tbi=180$ and
$\tbii=0\dots 200$. The 
rows are labelled by the corresponding
bulk minimal model; nodes are then labelled by the boundary condition
within that model. The highlighted flows, labelled in bold along the 
top of the figure, correspond to the values of $\tbii$ that were also
highlighted in figure \ref{gflows}d. Although some flow lines appear 
to cross, this is an artifact of the projection onto the page, and
does not occur in the full multidimensional
space of flows.
}
\end{array}
\]
\refstepcounter{figure}
\protect\label{firstflows}

These rules can be combined to understand the sequences of
$g$-function flows seen in figure~\ref{gflows} and further illustrated
in figures \ref{firstflows} and \ref{60}.
Consider $\tbi=180$,
$\tbii=50$, one of the highlighted flows
in figures \ref{gflows}d and \ref{firstflows}. 
Focussing on the part of the
flow beginning at $\mathcal{M}_8$, the initial $g$-function
value is close to that of the boundary condition $(1,4)$. Since
$\tbi>\ln(1/r)$ everywhere in this part of the flow, only 
$\tbii$ has an effect on the subsequent trajectory.
The centre of the $L(\te)$ plateau associated with the boundary
condition $(1,4)$ at $\mathcal{M}_8$ is at $\te=30$, and 
since $\tbii>30$ 
the flow within $\mathcal{M}_8$ is to $(3,1)$. Then as the bulk
theory flows from $\mathcal{M}_8$ to $\mathcal{M}_7$, the boundary
condition flows to $(1,3)$. The centre of the associated $L(\te)$
plateau is then at $\te=60$, and since $\tbii<60$, the flow within
$\mathcal{M}_7$ is then towards $(3,1)$, and so on. Repeating this
exercise for other values of $\tbii$ leads to the set of flows 
illustrated in figure~\ref{firstflows}, the second highlighted 
flow of which corresponds to $\tbii=50$.
Note that the pure-boundary flows
within $\CM_5$ on figure \ref{firstflows}
match the flows on the bottom
row of figure \ref{m=5}.

\[
\begin{array}{c}
\includegraphics[width=0.88\linewidth]{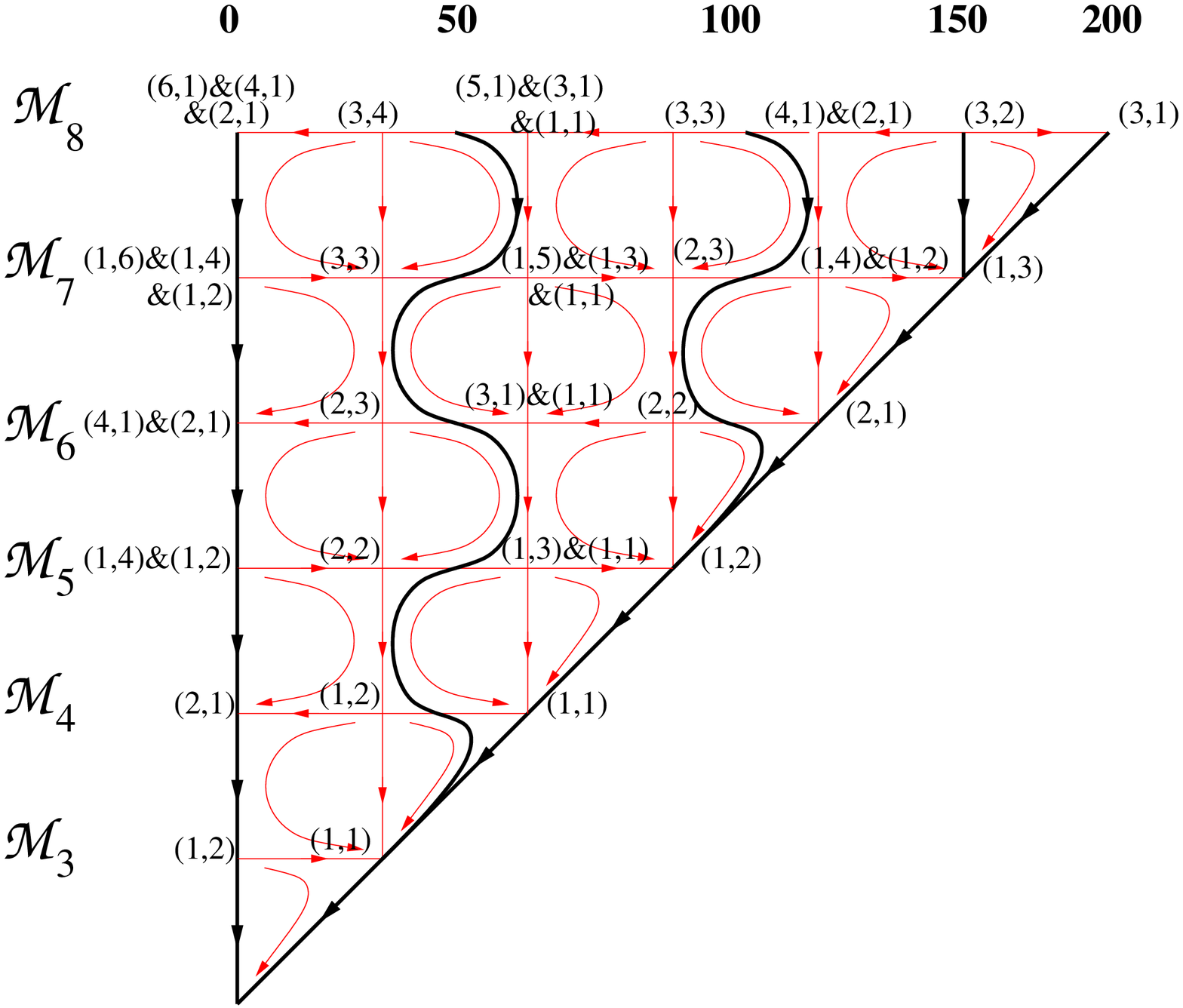}
\\[5pt]
\parbox{0.95\linewidth}{
\small Figure \protect\ref{60}: A depiction of the flows 
corresponding
to figure \ref{gflows}b, for which $\tbi=60$ and
$\tbii=0\dots 200$. Other labelling is as on figure
\ref{firstflows}.
}
\end{array}
\]
\refstepcounter{figure}
\protect\label{60}

We remarked while discussing
figure \ref{gflows} that the $g$-function plots
stabilise as $\tbi$ increases. The same feature can be seen on
comparing figure \ref{firstflows} with
the equivalent diagram for $\tbi=60$, figure \ref{60}.
For $\CM_3$ and $\CM_4$ the
boundary conditions and flows appearing in figure~\ref{60} match
those in figure \ref{firstflows}, but for the higher minimal
models the boundary conditions appearing are different. This
corresponds to how the plots in figures \ref{gflows}b and
\ref{gflows}d coincide for $\ln r>-60$.

\resection{Exact $g$-functions for $\CM A^{(+)}_m$}
\label{gMAplus}
Finally, we are ready to obtain the effective
equations which govern the $g$-function flows for the boundary
versions of the bulk interpolating
theories $\CM A^{(+)}_m$.
These theories
can be thought of as bulk perturbations of
the minimal models $\CM_m$ by their $\phi_{13}$ operators, with the
sign of the perturbation chosen so that the infrared limit of the
model is $\CM_{m-1}$, the next minimal model down. Our predictions for
the associated
boundary flows can be read from the results of the last
section, but to obtain exact equations, we need to take a 
careful limit of the $g$-function formula in parallel with the
double-scaling limit of the bulk TBA equations.

To fix notations we first review the situation for the bulk TBA. The 
TBA system proposed in \cite{Zamolodchikov:1991vx} for $\CM A^{(+)}_m$
involves $m-2$ pseudoenergies $\epsilon_1\dots\epsilon_{m-2}$, coupled
together by the following system of TBA equations:
\begin{align}
\epsilon_1(\theta)&=\frac{1}{2}\hat{r}\,e^\theta
-\int_{\RR}\phi(\theta-\theta')L_2(\theta')\,d\theta' \nn\\
\epsilon_a(\theta)&=
-\int_{\RR}\phi(\theta-\theta')
(L_{a-1}(\theta')+L_{a{+}1}(\theta'))\,d\theta'\qquad a=2\dots m{-}3 \nn\\
\epsilon_{m-2}(\theta)&=\frac{1}{2}\hat{r}\,e^{-\theta}
-\int_{\RR}\phi(\theta-\theta')L_{m-1}(\theta')\,d\theta' 
\label{bulkmasslessTBA}
\end{align}
where $L_a(\theta)=\ln(1+e^{-\epsilon_a(\theta)})$\,,
$\phi(\theta)=1/(2\pi\cosh(\theta))$ is as in (\ref{phidef}), and
$\hat{r}$ sets the crossover scale. The effective central charge
is then
\beq
c_{\rm eff}(\hat{r})=\frac{3\hat{r}}{2\pi^2}\int_{\RR}
(e^{\theta}L_1(\theta)+
e^{-\theta}L_{m-2}(\theta))\,d\theta
\eeq
and as $\ln(\hat{r})$ increases through a region of size of order 1
about the origin, $c_{\rm eff}(\hat{r})$ moves from $c_m$ to
$c_{m-1}$, consistent with the corresponding bulk flow being from $\CM_m$ 
to $\CM_{m-1}$.

These same equations emerge from a double-scaling limit of
the staircase model if we set
\beq
\ln(r) = -(m{-}3)\tz/2 + \ln(\hat{r})
\eeq
and then take the limit $\tz\to\infty$ keeping $\ln(\hat{r})$ 
finite, before finally allowing $\ln(\hat{r})$ to vary from
$-\infty$ to $+\infty$. In particular, the pseudoenergies
$\epsilon_a(\theta)$ are recovered from the staircase pseudoenergy
$\epsilon(\theta)$ in the $\theta_0\to\infty$ limit by setting
\beq
\epsilon_a(\theta)= \epsilon(\theta+(m{-}1{-}2a)\tz/2)\,,\qquad
a=1\dots m{-}2
\label{epsadef}
\eeq
and only allowing $\theta$ to vary over the full real line
after the limit has been taken.

It turns out that in this limit the staircase $g$-function formulae 
can be rewritten in terms of the limiting pseudoenergies
$\epsilon_a(\theta)$ and various constants which can be calculated in
terms of the plateau values of the staircase pseudoenergies. 
We start with the infinite series part of the $g$-function,
(\ref{lng0}). Crucial to the analysis 
is the double-bump shape of the kernel $\phi_S(\theta)$, shown in 
figure~\ref{phiSfig}, which causes each multiple integral contributing
to the sum to localise onto a collection of subregions of
$\RR^n$. In each of these subregions, the staircase pseudoenergy is
either constant, or else is uniformly well-approximated by one of the
interpolating-flow pseudoenergies $\epsilon_a(\theta)$. Rewriting the
formula for $\ln\go$ in terms of these constants and functions leads
to the effective equations which govern the $g$-function flow in $\CM
A^{(+)}_m$\,.

More precisely, since
$\phi_S(\theta)=\phi(\theta{-}\tz)+\phi(\theta{+}\tz)$, each term
$\phi_S(\theta_1+\theta_2)\phi_S(\theta_2-\theta_3)\cdots
\phi_S(\theta_n-\theta_1)$
in the sum in (\ref{lng0}) can be expanded as sum of $2^n$
terms of the form
\beq
\phi(\te_1+\te_2-\alpha_1\tz)\phi(\te_2-\te_3-\alpha_2\tz)\cdots
\phi(\te_n-\te_1-\alpha_n\tz)
\eeq
where each $\alpha_k=\pm1$. The decay properties of $\phi(\te)$ mean that
it is only non-zero for $\te\approx0$, so for the above term to be
non-zero as $\tz\to\infty$ we require
\begin{align}
 \te_1+\te_2&\approx\alpha_1\tz\nn\\
\te_2-\te_3&\approx\alpha_2\tz\nn\\[2pt]
\vdots&\nn\\[4pt]
\te_n-\te_1&\approx\alpha_n\tz
\end{align}
from which it follows that $\te_k\approx\overline\te_k$, $k=1\dots n$,
where
\begin{equation}
 \left(\begin{array}{c}
\overline\te_2\\
\overline\te_3\\\cdot\\\cdot\\\overline\te_n\\\overline\te_1
 \end{array}\right)
=
\half\tz\left(\begin{array}{cccccc}
1&1&1&1&\cdots&1\\
1&-1&1&1&\cdots&1\\
1&-1&-1&1&\cdots&1\\
\cdot&\cdot&\cdot&\cdot&&\cdot\\
\cdot&\cdot&\cdot&\cdot&&\cdot\\
1&-1&-1&-1&\cdots&-1
\end{array}\right)
\left(\begin{array}{c}
\alpha_1\\
\alpha_2\\
\cdot\\
\cdot\\
\cdot\\
\alpha_n
\end{array}
\right).
\label{matrix}
\end{equation}
and the integral over $\RR^n$ has indeed localised, to a set of $2^n$
regions of size of order one as $\tz\to\infty$, which become
infinitely separated in this limit.
The coordinates $\overline\te_k$ of the
centres of these regions are either all even multiples
of $\tz/2$, or all odd multiples of $\tz/2$, depending on whether $n$
is even or odd.

For each region of integration we must consider the behaviours
of the `measure factors' $1/(1+e^{\epsilon(\theta_k)})$, $k=1\dots n$.
As for the function $L(\theta)$ discussed above, these factors exhibit
a series of plateaux interleaved by transition regions, at
$\theta\approx\theta_a\equiv (m{-}1{-}2a)\tz/2$, 
$a=1\dots m{-}2$. Within these
transition regions, the measure factors
are well-approximated in the $\tz\to\infty$ limit by the
functions $1/(1+e^{\epsilon_a})$, by (\ref{epsadef}). In between these
regions the measure factors are approximately constant, and can be
expressed in terms of the numbers $y_a|_{\CM_m}=x_a|_{\CM_{m-1}}$,
$a=1\dots m{-}1$. Taking these considerations into account,
the terms in the sum in (\ref{lng0}) fall into two categories:
\begin{enumerate}[(a)]
\item If $m+n$ is odd, every $\overline\te_k$ satisfying 
$|\overline\te_k|\le
(m{-}3)\te/2$ lies in a transition region for $\epsilon(\te)$, so that
its measure factor remains nontrivial even after the
$\theta_0\to\infty$ limit has been taken.  We denote the
part of $\ln g_0$ consisting of these these terms by $\ln
\goA(\hat{r})$.
\item If $m+n$ is even, every $\overline\te_k$ lies inside a
plateau of $\epsilon(\te)$ after the $\te_0\to\infty$ limit has been
taken, so that the corresponding measure factor becomes constant. We 
denote the ($\hat{r}$-independent) part of $\ln g_0$ consisting 
of these these terms by $\ln \goB$.
\end{enumerate} 

For the $(a)$ terms, the values of $\epsilon(\te\approx\overline\te_k)$ 
vary as $\hat{r}$
varies, and only reach plateau values in the UV and IR limits, these
values being $x_{k+1}|_{\CM_m}$ in the UV and $y_k|_{\CM_{m-1}}$ in the
IR.
Rewriting the formulae in each subregion of integration in terms of
the limiting pseudoenergies $\ep_a(\theta)$ using (\ref{epsadef}) and
shifting the integration variables to remove
all appearances of $\tz$, we found that
$\ln \goA$ can be rewritten in the $\tz\to\infty$,
$\hat{r}$ finite limit as
\beq
\!\ln \goA(\hat{r})\,= 
\!\!\!\sum_{n\ge 1\atop m+n~{\rm odd}}\!\!
\frac{1}{2n}\int_{\RR^n}\!\!\mbox{antiTr}
\bigl(\Pi_{i=1}^nA(\te_i) d\te_i\bigr)
\phi(\te_1{-}\te_2)\phi(\te_2{-}\te_3)\cdots
\phi(\te_{n-1}{-}\te_n)\phi(\te_n{+}\te_1)
\label{goA}
\eeq
where the $(m-2)\times(m-2)$ matrix $A(\te)$ is given by
\beq
A(\te)=
\left(
\begin{array}{cccccccc}
0&\frac{1}{1+e^{\epsilon_1(\theta)}}&0&0&\cdots&0&0&0\\
\frac{1}{1+e^{\epsilon_2(\theta)}}&0&
\frac{1}{1+e^{\epsilon_2(\theta)}}&0&\cdots&0&0&0\\
0&\frac{1}{1+e^{\epsilon_3(\theta)}}&0&
\frac{1}{1+e^{\epsilon_3(\theta)}}&\cdots&0&0&0\\[7pt]
\vdots&\vdots&\vdots&\vdots&&\vdots&\vdots&\vdots\\[7pt]
0&0&0&0&\cdots&
\frac{1}{1+e^{\epsilon_{m-3}(\theta)}}&0&
\frac{1}{1+e^{\epsilon_{m-3}(\theta)}}\\
0&0&0&0&\cdots&0&\frac{1}{1+e^{\epsilon_{m-2}(\te)}}&0
\end{array}\right)
\label{Ai}
\eeq
and $\mbox{antiTr}\bigl(K\bigr)$,
the anti-trace of an $M\times M$ matrix $K$, is defined as 
the sum of its anti-diagonal elements, or equivalently
\beq
\mbox{antiTr}\bigl(K\bigr)=\mbox{Tr}(KJ)~,
\quad \mbox{where}~ J_{ij}=\delta_{i,M+1-j}\,.
\eeq

The measure factors for the (b) terms are by contrast
constant throughout the relevant integration subregions. They
can therefore be pulled outside their
integrals, leaving only the various factors of
$\phi(\te)$. This leads to the following expression for 
$\ln \goB$\,:
\begin{align}
\ln \goB\,&=
\!\!\!\sum_{n\ge 1\atop m+n~{\rm even}}
\frac{1}{2n}\mbox{antiTr}\bigl(B^n\bigr)\!
\int_{\RR^n}\!\!\phi(\te_1{-}\te_2)\phi(\te_2{-}\te_3)
\cdots\phi(\te_{n-1}{-}\te_n)\phi(\te_n{+}\te_1)
\,d\theta_1\!\cdots d\te_n\nn \\
&=\!\!\!
\sum_{n\ge 1\atop m+n~{\rm even}}
\frac{1}{n2^{n+2}}\,\mbox{antiTr}\left(B^n\right)
\,\,=\!\!
\sum_{n\ge 1\atop m+n~{\rm even}}
\frac{1}{n2^{n+2}}\,\mbox{Tr}\left(B^nJ\right)
\label{g0b}
\end{align}
where the tridiagonal
$(m-3)\times(m-3)$ matrix $B$ is equal to the limit as $\tz\to\infty$ of
\[\left(\begin{array}{cccccccc}
0&\frac{1}{1+e^{\epsilon\left(\frac{m-4}{2}\tz\right)}}&0&\cdots&0&0&0\\
\frac{1}{1+e^{\epsilon\left(\frac{m-6}{2}\tz\right)}}&0&
\frac{1}{1+e^{\epsilon\left(\frac{m-6}{2}\tz\right)}}&\cdots&0&0&0\\
0&\frac{1}{1+e^{\epsilon\left(\frac{m-8}{2}\tz\right)}}&0&
\cdots&0&0&0\\[7pt]
\vdots&\vdots&\vdots&&\vdots&\vdots&\vdots\\[7pt]
0&0&0&\cdots&\frac{1}{1+e^{\epsilon\left(-\frac{m-6}{2}\tz\right)}}&0
&\frac{1}{1+e^{\epsilon\left(-\frac{m-6}{2}\tz\right)}}\\
0&0&0&\cdots&0&\frac{1}{1+e^{\epsilon\left(-\frac{m-4}{2}\tz\right)}}&0
\end{array}\right)
\]
The explicit form of this matrix can be found using the $L(\te)$ plateau 
values (\ref{Ls}) and is
\[
\left(\begin{array}{ccccccc}
0&1-\frac{\sin^2\frac{\pi}{m}}{\sin^2\frac{2\pi}{m}}&0&\cdots&0&0&0\\
1-\frac{\sin^2\frac{\pi}{m}}{\sin^2\frac{3\pi}{m}}&0&
1-\frac{\sin^2\frac{\pi}{m}}{\sin^2\frac{3\pi}{m}}&\cdots&0&0&0\\
0&1-\frac{\sin^2\frac{\pi}{m}}{\sin^2\frac{4\pi}{m}}&0&
\cdots&0&0&0\\[7pt]
\vdots&\vdots&\vdots&&\vdots&\vdots&\vdots\\[7pt]
0&0&0&\cdots&1-\frac{\sin^2\frac{\pi}{m}}{\sin^2\frac{(m-3)\pi}{m}}
&0&1-\frac{\sin^2\frac{\pi}{m}}{\sin^2\frac{(m-3)\pi}{m}}\\
0&0&0&\cdots&0&1-\frac{\sin^2\frac{\pi}{m}}{\sin^2\frac{(m-2)\pi}{m}}&0
\end{array}\right)
\]
or, more concisely,
\beq
B_{ab}=l_{ab}\,x_{a+1}/(1{+}x_{a+1}) 
\eeq
where 
$x_a=x_a|_{\CM_{m-1}}=\frac{\sin^2(\pi a/m)}{\sin^2(\pi/m)}-1$\,,
and
$l_{ab}$ is the incidence matrix of the $A_{m-3}$ Dynkin diagram.
Note that $B$ is the transpose of a matrix which arises in the
analysis of small fluctuations about stationary solutions of an
associated Y-system \cite{Zamolodchikov:1991vx} and has 
eigenvalues
\beq
\lambda_k=2\cos\bigl(\frac{\pi k}{m}\bigr)\,,\qquad k=2,3\dots m{-}2\,.
\eeq
For later use we note that $[B,J]=0$, and furthermore that
the eigenvector $\psi_k$ of $B$ corresponding to the
eigenvalue $\lambda_k$ satisfies 
\beq
J\psi_k=(-1)^k\psi_k\,.
\label{Jparity}
\eeq

As explained in appendix A, this information is enough to evaluate
(\ref{g0b}) in closed form, with the result
\begin{numcases}{\ln \goB|_{\CM A^{(+)}_m} = }
\displaystyle
 ~\ln \left(\left(\frac{4}{m}\right)^{\frac{1}{4}}\!
\frac{\sin\frac{(m-1)\pi}{2m}}
{\sqrt{\sin\frac{\pi}{m}}}\right)
&\mbox{~~for $m$ odd}
\label{g0bodd}
\\[15pt]
 ~\ln \left(\left(\frac{2}{m}\right)^{\frac{1}{4}}\!\!
\frac{1}{\sqrt{\sin\frac{\pi}{m}}}\right)
&\mbox{~~for $m$ even}
\label{g0beven}.
\end{numcases}
A missing piece of the staircase discussion from the previous section
can now be filled in,
namely the formulae 
(\ref{lng0fodd}) and (\ref{lng0feven})
for the value of $\ln \go|_{\CM_m}$ when the bulk 
staircase theory is in
the vicinity of $\CM_m$. In terms of the limiting $g$-function
equations for $\CM A^{(+)}_m$, this number is equal to the UV limit of
$\ln \goA(\hat{r}) + \ln \goB$ as $\hat{r}\to 0$.
Considering the limiting forms of the matrix $A(\te)$, given by
(\ref{Ai}), as $\hat r\to 0$, it is
straightforwardly seen that
\beq
\ln \goA(0)|_{\CM A^{(+)}_m}=\ln \goB|_{\CM A^{(+)}_{m+1}}
\label{gUV}
\eeq
and so
\beq
\ln \go|_{\CM_m} = 
\ln \goB|_{\CM A^{(+)}_{m+1}} +
\ln \goB|_{\CM A^{(+)}_m} 
\eeq
\nobreak
which, via (\ref{g0bodd}) and (\ref{g0beven}),
leads immediately to (\ref{lng0fodd}) and
(\ref{lng0feven}).

The remainder of the exact $g$-function,
$\ln\gb=\ln\gi+\ln\gii+\ln\giii$, is
more straightforward to analyse. 
In the limit $\{\tz\to\infty,\,\ln\hat r~\mbox{finite}\}$,
the behaviours of the first two terms, $\ln\gi$ and $\ln\gii$, depend
on whether $m$ is even or odd.

The presence of $\phi(\te)$ and $\delta(\te)$ in the expression
(\ref{lngb1}) for $\ln\gi$ 
means that this term is determined by the staircase pseudoenergy
close to $\te=0$. When $m$ is odd this remains nontrivial as the 
double-scaling limit is taken and, using
the pseudoenergies defined in (\ref{epsadef}), $\ln\gi$ becomes
\beq
\ln\gi
= -\half\int_{\RR}{d\te\left(\phi(\te)+
\half\delta(\theta)\right)\!\ln(1+e^{-\epsilon_{\frac{m-1}{2}}(\te)})}\,,
\eeq
which has UV and IR limits given by (\ref{lngb1odd}) at $\CM_m$ and
(\ref{lngb1even}) at $\CM_{m-1}$ respectively.  When $m$ is even, the 
staircase pseudoenergy instead becomes constant near to
$\te=0$, giving
\beq
\ln\gi=-\half\ln(1+y_{m/2}|\phup_{\CM_{m}})
\eeq
matching (\ref{lngb1even}) at $\CM_m$ and also (\ref{lngb1odd}) at
$\CM_{m-1}$.

In contrast, the formula 
(\ref{lngb2}) for $\ln\gii$ involves
$\phi_{(\frac{3}{4})}(\te\pm\half\tz)$ 
and $\phi(2\theta\pm\tz)$,
meaning that $\ln\gii$ is determined by
the behaviour of 
$L(\te)$ close to $\te=\pm\tz/2$.  For $m$ odd, the limiting 
form of $L(\theta\approx\pm\tz/2)$ is constant, and
\beq
\ln\gii=-\half\ln(1+y_{\frac{m-1}{2}}|_{\CM_{m}})
\eeq
as in (\ref{lngb2odd}) at $\CM_m$ and (\ref{lngb2even}) at $\CM_{m-1}$.
When $m$ is even, $L(\theta\approx\pm\tz/2)$ 
remains nontrivial and in the limit
\beq
\ln\gii=\int_{\RR}{d\te\bigl(\phi_{(\frac{3}{4})}(\te) -
\phi(2\theta)\bigr)\!\ln(1+e^{-\epsilon_{\frac{m-2}{2}}(\te)})}\,,
\eeq
with UV and IR limits given by (\ref{lngb2even}) at $\CM_m$
and (\ref{lngb2odd}) at $\CM_{m-1}$ respectively.

To allow the last part of $\ln\gb$, $\ln\giii$, 
to retain a non-trivial $\hat r$-dependence in
the limit, we pick two integers $a_1$ and $a_2$ with 
$0\le a_i\le m{-}1$, 
write the boundary parameters $\tbi$ and $\tbii$ 
as
\beq
\theta_{b_i}=\half(m{-}1{-}2a_i)\tz +\hat\theta_{b_i}
\eeq
for $i=1,2$, and then
take the $\tz\to\infty$ limit
keeping $\hat\theta_{b1}$ 
and $\hat\theta_{b2}$ finite.
Given the specification (\ref{epsadef}) of the effective
pseudoenergies $\ep_a(\theta)$, 
for $1\le a_i\le m{-}2$
the staircase expression
(\ref{lngb3}) for $\ln\giii$ then reduces to
\begin{align}
\ln g\phup_{b3\,\,a_1a_2}(\hat r,\htbi,\htbii) &=
\half\sum_{i=1}^2\int_{\RR}{d\te\bigl(\phi(\te-\hat\te_{b_i})+
\phi(\te+\hat\te_{b_i})
\bigr)\!\ln(1+e^{-\ep_{a_i}(\te)})} \nn\\
&=
\sum_{i=1}^2\int_{\RR}{d\te\,\phi(\te-\hat\te_{b_i})
\ln(1+e^{-\ep_{a_i}(\te)})}
\label{gbiii}
\end{align}
where the symmetry $\epsilon_a(\te)=\epsilon_{m-1-a}(-\te)$ 
of the ground-state pseudoenergies for $\CM A^{(+)}_m$
was used in going from the first line to the second.
If either $a_i$ is equal to $0$ or $m{-}1$, then the
staircase pseudoenergy diverges in the region of $\te_{b_i}$, and the
corresponding term in (\ref{gbiii}) is zero in the $\CM A^{(+)}_m$
limit.

To summarize the results of this section, our final expressions 
for the two-parameter families of exact 
$g$-functions for $\CM A^{(+)}_m$, indexed by a pair of integers $a_1$
and $a_2$ and expressed in terms of
the rescaled variables $\hat r$, $\htbi$ and 
$\htbii$, are as follows:
\vskip -25pt
\begin{align}
\intertext{ $\bullet$ $m$ odd:}
\ln g_{a_1a_2}(\hat{r},\htbi,\htbii) &=
\ln\!\left(\left(\frac{4}{m}\right)^{\frac{1}{4}}\!\!\!
\sqrt{\sin\frac{\pi}{m}}\right)-\half\int_{\RR}{d\te\left(\phi(\te)+
\half\delta(\theta)\right)\!\ln(1+e^{-\epsilon_{\frac{m-1}{2}}(\te)})}
\nn\\[5pt]
&\quad\qquad +\ln\goA(\hat{r})
+\ln g\phup_{b3\,\,a_1a_2}(\hat r,\htbi,\htbii)\,;
\label{gmodd}\\
\intertext{ $\bullet$ $m$ even:}
\ln g_{a_1a_2}(\hat{r},\htbi,\htbii) &=
\ln\!\left(\left(\frac{2}{m}\right)^{\frac{1}{4}}\!\!\!
\sqrt{\sin\frac{\pi}{m}}\right)+\int_{\RR}{d\te\Bigl(\phi_{(\frac{3}{4})}(\te)
- \phi(2\theta)\Bigr)\ln(1+e^{-\epsilon_{\frac{m-2}{2}}(\te)})}
\nn\\[5pt]
&\quad\qquad 
+\ln\goA(\hat{r})
+\ln g\phup_{b3\,\,a_1a_2}(\hat
r,\htbi,\htbii)\,.
\label{gmeven}
\end{align}
In both cases $\goA$ is given by (\ref{goA}), the
sum over $n$ running through even integers for $m$ odd, and 
odd integers for $m$ even, with the pseudoenergies involved
solving the bulk
$\CM A^{(+)}_m$ TBA system (\ref{bulkmasslessTBA}). The term
$g\phup_{b3\,\,a_1a_2}$ is as defined in (\ref{gbiii}). The
constant terms result from adding $\ln\goB$, given by
(\ref{g0bodd}) or (\ref{g0beven}), to $\ln\gii$ for $m$ odd, and to
$\ln\gi$ for $m$ even. The remaining integral term is $\ln\gi$
for $m$ odd and $\ln\gii$ for $m$ even. Formally setting one or both
of $a_1$ and $a_2$ equal to $0$ or $m{-}1$, as discussed after
(\ref{gbiii}), incorporates the
limiting one- and zero- parameter families
of flows found by deleting the $\htbi$ and/or
$\htbii$ dependent parts of $\ln\giii$. If both are
deleted so that $\ln\giii$ is identically zero, 
then by the results of the last section the
$g$-function flow should be from $g(m,1,1)$ at $\hat r=0$ to
$g(m{-}1,1,1)$ as $\hat r\to\infty$. This can be checked directly.
Simplest is the UV limit. For $m$ odd, 
using (\ref{lngb1odd})
and (\ref{gUV}),
the first three terms on the RHS of (\ref{gmodd}) tend to
\beq
\ln\left(\left(\frac{4}{m}\right)^{\frac{1}{4}}
\!\!\sqrt{\sin{\frac{\pi}{m}}}\right)
-\half\ln\left(\frac{1}{\sin^2\left(\frac{\pi}{m+1}\right)}\right)
+\ln\left(\left(\frac{2}{m+1}\right)^{\frac{1}{4}}
\frac{1}{\sqrt{\sin{\frac{\pi}{m+1}}}}\right)
\eeq
which is equal to $g(m,1,1)$\,.
Similarly, for $m$ even, (\ref{lngb2even}) and (\ref{gUV})
imply that the first three terms on the RHS of
(\ref{gmeven}) become
\beq
\ln\left(\left(\frac{2}{m}\right)^{\!\frac{1}{4}}\!\!
\sqrt{\sin{\frac{\pi}{m}}}\right)
-
\half\ln\left(\frac{\sin^2\left(\frac{\pi m}{2(m+1)}\right)}%
{\sin^2\left(\frac{\pi}{m+1}\right)}\right)
+
\ln\left(\left(\frac{4}{m{+}1}\right)^{\!\frac{1}{4}}
\frac{\,\sin\left(\frac{m\pi}{2(m+1)}\right)}%
{\sqrt{\sin{\frac{\pi}{m+1}}}}\right)
\eeq
\nobreak
and the value of $g(m,1,1)$ is again reproduced.
\goodbreak

In the $\hat r\to\infty$, IR, limit
the matrix $A(\te_i)$ defined in (\ref{Ai}) again becomes independent
of $\te$ in the central region $\te\approx 0$ where the integrals in
(\ref{goA}) have their support, and tends to the following 
$(m{-}2)\times (m{-}2)$ matrix:
\beq
A_{\rm IR}=
\left(\begin{array}{cccccccc}
0&0&0&\cdots&0&0&0\\
1-\frac{\sin^2\frac{\pi}{m-1}}{\sin^2\frac{2\pi}{m-1}}&0
&1-\frac{\sin^2\frac{\pi}{m-1}}{\sin^2\frac{2\pi}{m-1}}&\cdots&0&0&0\\
0&1-\frac{\sin^2\frac{\pi}{m-1}}{\sin^2\frac{3\pi}{m-1}}
&0&\cdots&0&0&0\\
&&&&&&\\
\vdots&\vdots&\vdots&&\vdots&\vdots&\vdots\\
&&&&&&\\
0&0&0&\cdots&1-\frac{\sin^2\frac{\pi}{m-1}}{\sin^2\frac{(m-3)\pi}{m-1}}
&0&1-\frac{\sin^2\frac{\pi}{m-1}}{\sin^2\frac{(m-3)\pi}{m-1}}\\
0&0&0&\,\cdots\,&0&~~0~~&0
\end{array}\right).\eeq
In terms of $A_{\rm IR}$\,,
\beq
\ln\goA(\infty)=\sum_{n\ge 1\atop m+n~{\rm odd}}
\frac{1}{n2^{n+2}}\,\mbox{Tr}\left((A_{\rm IR})^nJ\right).
\eeq
To evaluate this sum, observe that the central $(m{-}4)\times(m{-}4)$
sub-matrix of $A_{\rm IR}$ is the matrix $B$ for $\CM A^{(+)}_{m-1}$.
Furthermore, the zero values of all entries in the first and final
rows means that the central $(m{-}4)\times(m{-}4)$ sub-matrix of 
$(A_{\rm IR})^n$ is just 
$(B|_{\CM A^{(+)}_{m-1}})^n$\,, 
and the entries on the first and final rows are again zero.
Hence the traces and antitraces of $(A_{\rm IR})^n$ are equal to those
of $(B|_{\CM A^{(+)}_{m-1}})^n$\,, and 
\beq
\ln g_{0_A}(\infty)|_{\CM A^{(+)}_{m}}=\ln g_{0_B}|_{\CM A^{(+)}_{m-1}}
\eeq
allowing (\ref{g0bodd}) and (\ref{g0beven}) at $m{-}1$
to be used to compute
$\ln g_{0_A}(\infty)|_{\CM A^{(+)}_{m}}$\,.
For $m$ odd, using (\ref{lngb1even}) and
(\ref{g0beven}) at $m{-}1$,
the first three terms of (\ref{gmodd}) therefore become
\beq
\ln\left(\left(\frac{4}{m}\right)^{\frac{1}{4}}\!\!
\sqrt{\sin{\frac{\pi}{m}}}\right)
-\half\ln\left(\frac{1}{\sin^2\left(\frac{\pi}{m-1}\right)}\right)
+ \ln\left(\left(\frac{2}{m-1}\right)^{\frac{1}{4}}\!\!
\frac{1}{\sqrt{\sin{\frac{\pi}{m-1}}}}\right)
\nn\\
\eeq
which is indeed equal to $g(m{-}1,1,1)$. 
Similarly, for $m$ even, in the $\hat r\to\infty$ limit 
the first three terms of (\ref{gmeven}) are, by 
(\ref{lngb2odd}) and (\ref{g0bodd}) at $m{-}1$,
\beq
\ln\left(\left(\frac{2}{m}\right)^{\frac{1}{4}}\!\!
\sqrt{\sin{\frac{\pi}{m}}}\right)
-\half\ln\left(\frac{\sin^2\left(\frac{(m-2)\pi}{2(m-1)}
\right)}{\sin^2\left(\frac{\pi}{m-1}\right)}\right)
+\ln\left(\left(\frac{4}{m-1}\right)^{\frac{1}{4}}
\frac{\sin\left(\frac{(m-2)\pi}{2(m-1)}\right)}%
{\sqrt{\sin{\frac{\pi}{m-1}}}}\right)
\eeq
which again matches $g(m{-}1,1,1)$.

More generally, the full equations (\ref{gmodd}) or (\ref{gmeven})
predict a collection of two-parameter families of
flows, indexed by the two integers $a_1$ and $a_2$.
The relevant calculations have been carried out in the last section
and we won't repeat them here. Instead, in figure \ref{cubef} below we
show a typical family of flows,
where $r=a_1+1$ and $s=a_2+1$. 
All flows for $\htbi$ and $\htbii$ finite start in the far UV at
the $[x_r,x_s]$ boundary of $\CM_m$, a superposition of Cardy
boundaries given by the rule (\ref{gident}). If both $\htbi$ and
$\htbii$ are zero, the flow is directly downwards to the
$[y_{r-1},y_{s-1}]$ boundary of $\CM_{m-1}$, driven by the bulk
perturbation. Nonzero values of $\htbi$ and $\htbii$ 
correspond to the addition of boundary perturbations to the bulk
perturbation, and cause the
trajectory to visit other boundaries on its way from UV to IR, as can
be read from the figure.
If either of $a_1$ or $a_2$ is equal to $1$ or $m{-}2$, so that one or
both of $r$ and $s$ is equal to $2$ or $m{-}1$, the cube shown in
figure \ref{cubef} truncates, the equalities $x_1=y_1$ or
$x_{m-1}=y_{m-2}$ within $\CM_{m-1}$ causing one or two rows on the
bottom face of the cube to fuse together, resulting in the
flow patterns illustrated in
figure \ref{cubetrunc}. The same phenomenon was seen earlier for the 
staircase model, and is reflected in the diagonal lines of
flows running from top right to bottom left in figures \ref{firstflows} 
and \ref{60}.

\[
\begin{array}{c}
\\[-35pt]
\includegraphics[width=0.85\linewidth]{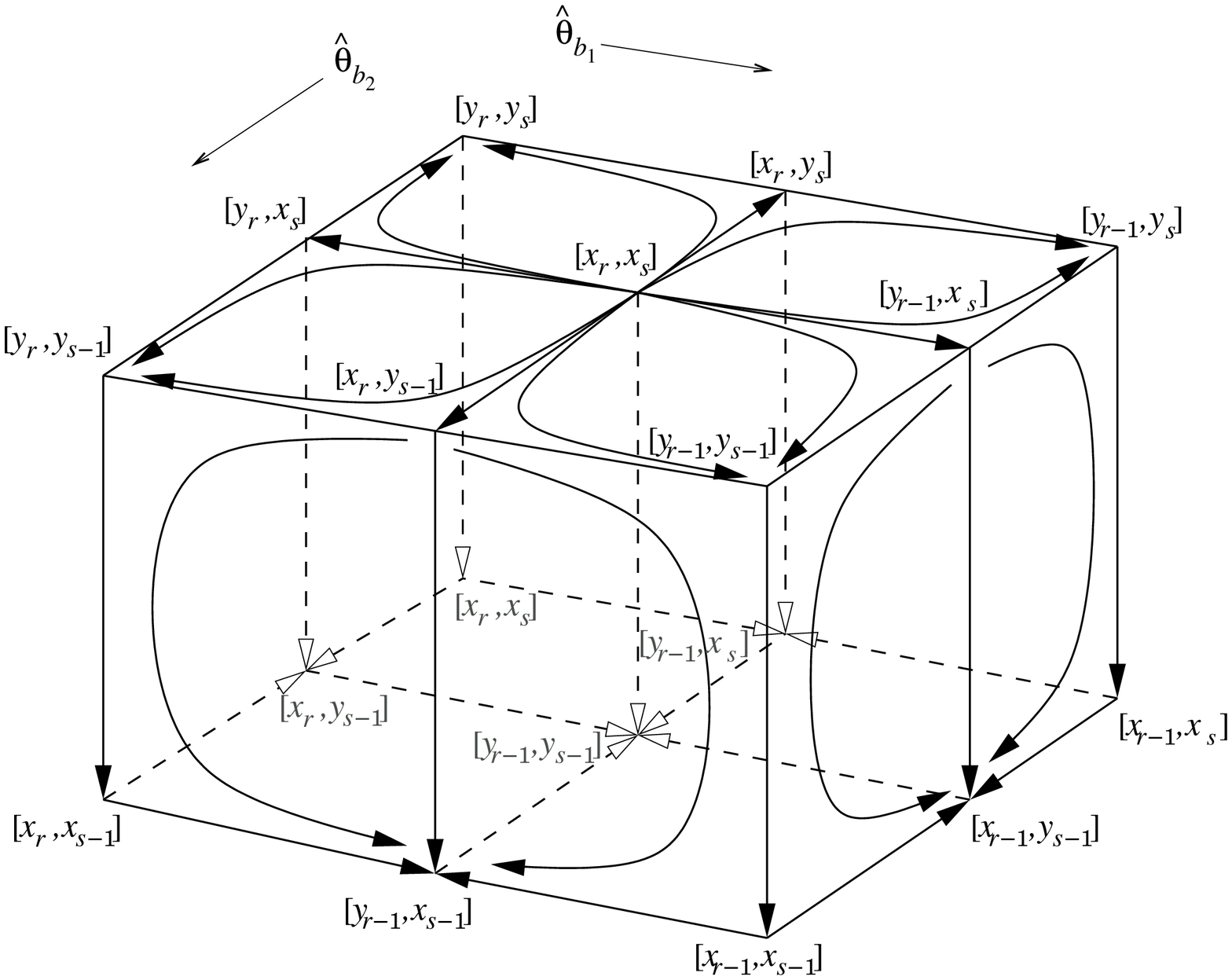}
\\[5pt]
\parbox{0.98\linewidth}{
\small Figure \protect\ref{cubef}: A cube of $\CM A^{(+)}_m$ flows, 
where $r=a_1{+}1$ and $s=a_2{+}1$.
Flows on the outer faces 
occur for $\htbi\to\pm\infty$ and/or $\htbii\to\pm\infty$;
those on the top face are within the minimal model $\CM_m$, and
those on the bottom are within $\CM_{m-1}$. 
The central vertical axis corresponds to $\htbi=\htbii=0$. 
Conformal boundary conditions at fixed points of the flows are 
labelled according to the scheme given in equation~(\ref{gident}).
}
\end{array}
\]
\refstepcounter{figure}
\protect\label{cubef}

\[
\begin{array}{c}
\\[-30pt]
\!\!\!\!
\!\!\!\!
\!\!\!\!
\!\!\!\!
\includegraphics[width=0.6\linewidth]{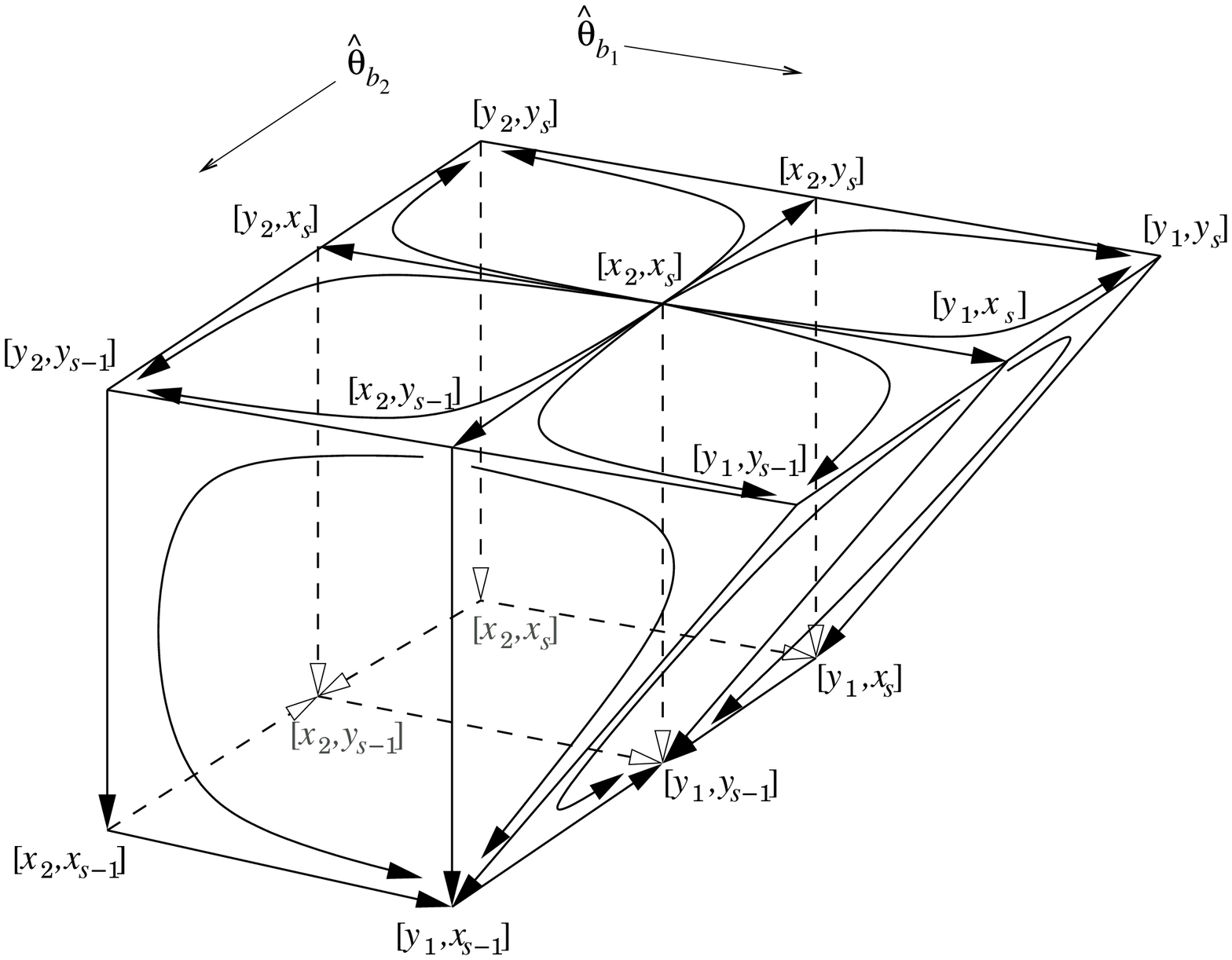}
\!\!\!\!
\!\!\!\!
\!\!\!\!
\!\!\!\!
\!\!
\includegraphics[width=0.6\linewidth]{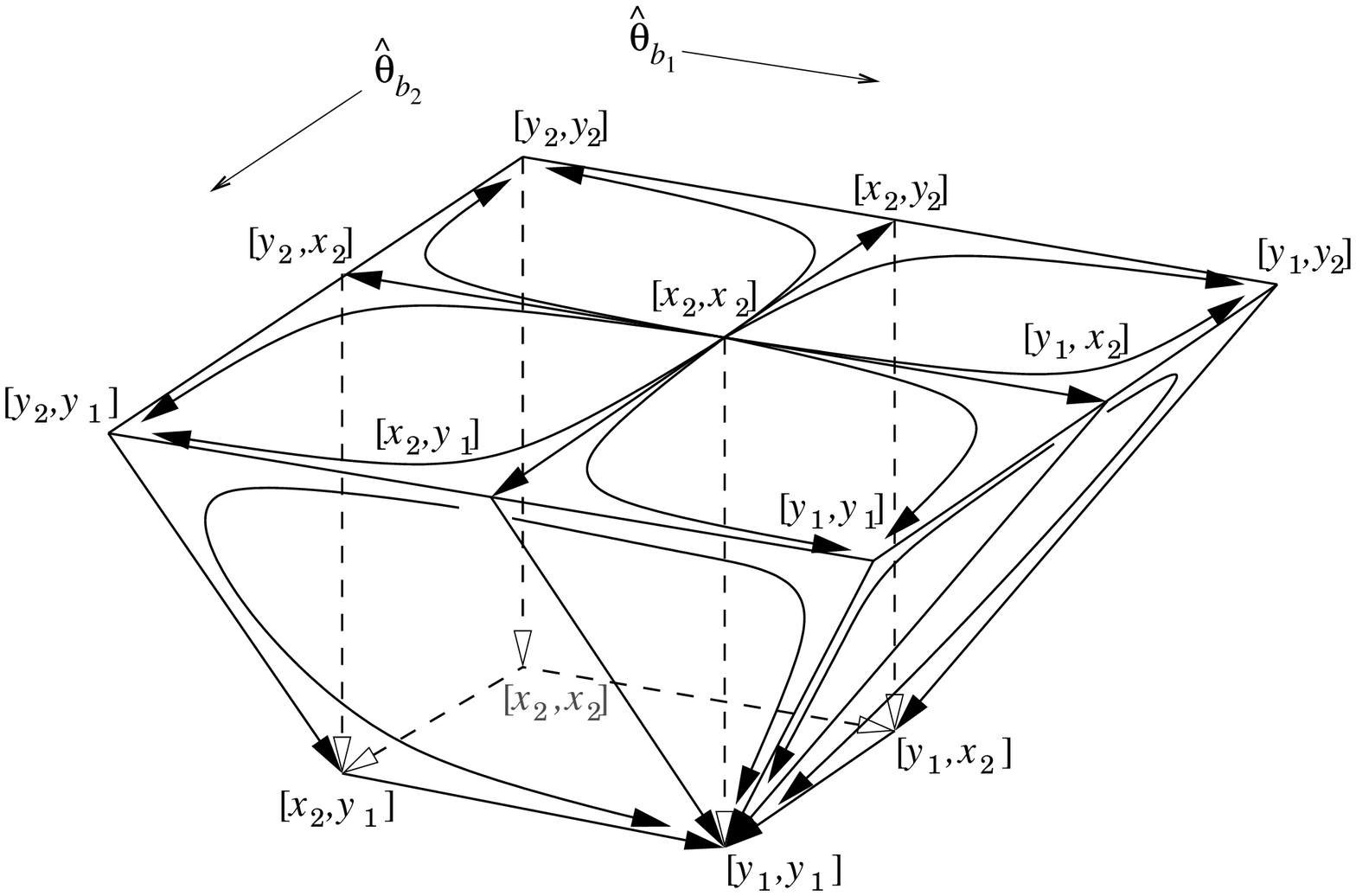}
\\[5pt]
\!\!\!\!\!\!\!\!\!\!\!\!\!\!\!\!\!\!\!\!\!\!\!
\parbox{0.98\linewidth}{
\small Figure \protect\ref{cubetrunc}: Truncations of the cube of 
flows for $r=2$ (left) and $r=s=2$ (right), corresponding to setting
either $a_1=1$ or $a_1=a_2=1$ in (\ref{gbiii}).
Other details are
as on figure \ref{cubef}.
}
\end{array}
\]
\refstepcounter{figure}
\protect\label{cubetrunc}


\resection{Conclusions}
In this paper we have shown how exact methods can be used to study
interpolating boundary flows in two-dimensional integrable models in
situations where both bulk and boundary are away from criticality. 
Our results for flows between minimal models
have confirmed and extended previous perturbative
studies. In addition we have shown how these flows can be embedded
into
larger manifolds of boundary integrability via the staircase model,
generalising the picture seen in the bulk. In fact the staircase 
description is remarkably economical -- the bulk S-matrix
(\ref{stairS}) and the boundary reflection factor (\ref{R}) (or
(\ref{Rn})) together encode not only all of the unitary $c<1$
minimal models, but also all of their Cardy boundary conditions and a
variety of their superpositions, once
fed into the general TBA and exact $g$-function machinery.

There are many directions for future work. Some perturbative checks 
of the exact equations for
the tricritical to critical Ising bulk and boundary flows were
undertaken in \cite{Dorey:2009vg}, but further tests higher up the
series would be valuable, as would a more detailed study of the
two-parameter boundary flows discussed in section \ref{sec3}, and
their extensions via defects to incorporate further parameters through
the more general reflection factor (\ref{Rn}).  
The formulae found in section~\ref{gMAplus}
describe, for the first time, exact off-critical $g$-functions
in situations where the underlying scattering theory is
non-diagonal, and it will be interesting to generalise the approach of
Pozsgay~\cite{Pozsgay:2010tv} to cover such cases. At the same time, 
many other multiparameter families of integrable models
with nontrivial intermediate scaling behaviours 
are now known, including generalisations of the staircase models 
\cite{Martins:1992ht,Dorey:1992bq,Martins:1992sx,Dorey:1992pj}, and
the Homogeneous Sine-Gordon (HSG) models 
\cite{FernandezPousa:1996hi,CastroAlvaredo:1999em,Dorey:2004qc}.
Hence there is plenty of scope to obtain more elaborate
exact $g$-function flows using the approach adopted in this paper. 
Finally, it is noteworthy how the embedding of non-diagonal bulk and 
boundary scattering theories within higher-dimensional manifolds of 
integrability achieved by the staircase and HSG models manages to
`abelianise' their TBA descriptions. It would be very interesting to
know how general this phenomenon is, and to understand it at a deeper
level. At the very least, it demonstrates once again that simple
exact S-matrices and reflection factors
can hide a great deal of internal structure.

\bigskip
\medskip
\noindent
{\bf Acknowledgements --}
We would like to thank Peter Bowcock, Ed Corrigan, Matthias Gaberdiel, 
Bal\'azs Pozsgay, Gabor Takacs and G\'erard Watts for interesting and 
helpful discussions about this project, and, especially, Chaiho Rim for
collaboration in its early stages.
PED thanks the Perimeter Institute and the Centro de Ciencias de 
Benasque Pedro Pascual for hospitality. The work was supported in part
by an STFC rolling grant, number ST/G000433/1, and by an STFC
studentship (RMW).

\appendix
\label{appA}
\appsection{Proofs of equations
(\ref{g0bodd}) and (\ref{g0beven})}
To derive (\ref{g0bodd}) and (\ref{g0beven}) from (\ref{g0b}), we 
first convert the final sum, over traces of products of powers of $B$
with $J$,
into a sum of traces of pure powers of
related matrices, after which the identity
\beq
\sum_{n=1}^{\infty}\frac{1}{n}\mbox{Tr}\,M^n
= -\mbox{Tr}\ln(I-M)
= -\ln\mbox{Det}(I-M)
\label{lntr}
\eeq
will allow the sum to be evaluated in terms of the eigenvalues of
these related matrices. There are two cases. 

\medskip

\noindent $\bullet$ If $m$ is odd, then the sum 
in (\ref{g0b}) is over odd powers of $B$. Shuffling indices and
using the symmetries of $B$ it can be checked that $B^nJ=(BJ)^n$ for
all odd $n$, and so 
\begin{eqnarray}
 \ln \goB&=&
\frac{1}{4}
\mbox{Tr}
\sum_{n\ge 1\atop n {\rm odd}}
\frac{1}{n}\left(\half BJ\right)^n\nn\\
&=&\frac{1}{8}
\ln\frac{\mbox{Det}\left(1+\half BJ\right)}{\mbox{Det}\left(1-\half
BJ\right)}\,.
\label{lnodd}
\end{eqnarray}
From (\ref{Jparity}), the
eigenvalues of $BJ$ are $(-1)^k\lambda_k$, which for $m$ odd are the
numbers $2\cos(2\pi/m)$, $2\cos(4\pi/m)$\,\dots
$2\cos((m{-}3)\pi/m)$, all with multiplicity two. Hence 
\begin{equation}
\frac{\mbox{Det}\left(1+\half BJ\right)}{\mbox{Det}\left(1-\half
BJ\right)}=
\frac{\cos^4\frac{\pi}{m}\cos^4\frac{2\pi}{m}\cdots\cos^4\frac{(m-3)\pi}{2m}}%
{\sin^4\frac{\pi}{m}\sin^4\frac{2\pi}{m}\cdots\sin^4\frac{(m-3)\pi}{2m}}\,.
\end{equation}
To evaluate this we call upon some well-known trigonometric
identities. For the numerator we use
\begin{equation}
\cos\frac{\pi}{n}\cos\frac{2\pi}{n}\cdots\cos\frac{(n-1)\pi}{2n}=
\frac{1}{2^{(n-1)/2}}\quad\mbox{for $n$ odd}
\label{trig0}
\end{equation}
and for the denominator 
\beq
\sin\frac{\pi}{n}\sin\frac{2\pi}{n}\cdots\sin\frac{(n-1)\pi}{n}=
\frac{n}{2^{n-1}}
\label{trig}
\eeq
which implies
\beq\sin^2\frac{\pi}{n}\sin^2\frac{2\pi}{n}\cdots\sin^2\frac{(n-1)\pi}{2n}
=\frac{n}{2^{n-1}}\quad\mbox{for $n$ odd.}
\label{trig2}\eeq
Using (\ref{trig0}) and (\ref{trig2}),
\beq
 \frac{\mbox{Det}\left(1+\half BJ\right)}%
{\mbox{Det}\left(1-\half BJ\right)}=
\frac{1}{m^2}\frac{\sin^4\frac{(m-1)\pi}{2m}}{\cos^4\frac{(m-1)\pi}{2m}}
=\left(\frac{4}{m}\frac{\sin^4\frac{(m-1)\pi}{2m}}%
{\sin^2\frac{\pi}{m}}\right)^2
\eeq
and so (\ref{lnodd}) is in agreement with (\ref{g0bodd}).

\medskip

\noindent $\bullet$ If $m$ is even, then the sum in (\ref{g0b}) is 
instead over even powers of $B$. This time we make use of the
identity, valid for $m$ even, that
\beq 
\mbox{Tr}(B^nJ)=\mbox{Tr}(B^n-B^{\prime n})
\label{traceident}
\eeq
where $B^{\prime}$ is the $(m-3)\times(m-3)$ matrix which only differs
from $B$ in the following elements:
\beq
B^{\prime}_{\frac{m-2}{2},\frac{m-4}{2}}=
B^{\prime}_{\frac{m-2}{2},\frac{m}{2}}=0.
\label{BPdef}
\eeq
To prove this identity we write the LHS as
\beq 
\mbox{Tr}(B^nJ)=
B_{i_1i_2}B_{i_2i_3}\cdots B_{i_{n-1}i_{n}}B_{i_{n}m-i_1-2}
\label{Banti}
\eeq
and note the following properties of $B$:
\beqa
B_{i,i-1}&=&B_{i,i+1}, \mbox{\quad all other entries $0$;}\label{B0}\\
B_{ij}&=&B_{m-2-i,m-2-j}\,.\label{B1}
\label{B2}
\eeqa
The first of these means that (\ref{Banti}) can be interpreted as a
weighted sum over all $n$-step paths on the $A_{m-3}$ Dynkin diagram which
start and finish at pairs of conjugate nodes
$i_1$ and
$m{-}2{-}i_1$, $i_1=1\dots m{-}3$,
and move by one link at each step.
Likewise $\mbox{Tr}(B^n)$ and $\mbox{Tr}(B'^n)$ are 
weighted sums over
$n$-step paths on the same Dynkin diagram, but which this time
start and finish at the same node $i_1$, where again $i_1$ is summed
from $1$ to $m{-}3$.

These observations imply that both sides of (\ref{Banti}) are zero for
$n$ odd, and so from now on we can take $n$ to be even (which is the
case of direct interest in the current context). 
Due again to (\ref{B0}), a term in $\mbox{Tr}(B^nJ)$ with
$i_1<\frac{m-2}{2}$ must include the element
$B_{\frac{m-2}{2}\frac{m}{2}}$, and a term with
$i_1>\frac{m-2}{2}$ must include the element
$B_{\frac{m-2}{2}\frac{m-4}{2}}$. 
Assume first that $i_1<\frac{m-2}{2}$, and consider
\beq
B_{i_1i_2}B_{i_2i_3}\cdots B_{i_{n-1}i_{n}}B_{i_{n},m-2-i_1}
\label{Bterm}
\eeq
for some particular $i_2\dots i_n$. Suppose
$B_{\frac{m-2}{2}\frac{m}{2}}$ appears for the final time at
$B_{i_pi_{p+1}}$, which means that
$B_{i_{p+1}i_{p+2}}=B_{\frac{m}{2}\frac{m+2}{2}}$. 
By (\ref{B1}), the value of (\ref{Bterm}) is unchanged if all
indices $i_q$ with $q>p$ are replaced by their conjugates
$m-2-i_q$. The resulting term appears in the
trace not of $B^nJ$, but of $B^n$. Similarly, every term in the
expansion of $\mbox{Tr}(B^nJ)$ with $i_1>\frac{m-2}{2}$ can be equated
with a term in the expansion of $\mbox{Tr}(B^n)$ with
$i_1>\frac{m-2}{2}$\,. Finally,
the terms in the expansions of $\mbox{Tr}(B^nJ)$ and $\mbox{Tr}(B^n)$
with $i_1=\frac{m-2}{2}$ are already equal, since $i_1$ is then equal
to $m-2-i_1$. Thus $\mbox{Tr}(B^nJ)$ is equal to the sum of the terms
in the trace of $B^n$ that include either
$B_{\frac{m-2}{2}\frac{m}{2}}$ or
$B_{\frac{m-2}{2}\frac{m-4}{2}}$ at least once. Now the trace of 
$B'^n$ as defined by (\ref{BPdef})
is equal to the trace of
$B^n$ minus the terms where $B_{\frac{m-2}{2}\frac{m}{2}}$ or
$B_{\frac{m-2}{2}\frac{m-4}{2}}$ appears at least once. Hence
$\mbox{Tr}(B^n-B^{\prime n})$ gives the required terms and
(\ref{traceident}) holds.

If $m$ is even we therefore have
\begin{eqnarray}
 \ln \goB&=&
\sum_{n>1\atop n~{\rm even}}
\frac{1}{n2^{n+2}}\mbox{Tr}\,(B^nJ)\nn \\
&=&\frac{1}{4}
\sum_{n>1\atop n~{\rm even}}
\frac{1}{n}\mbox{Tr}\left(\left(\half
B\right)^n-\left(\half B^{\prime}\right)^n\right)\nn\\
&=&\frac{1}{8}\ln
\frac{\mbox{Det}\left(I-\frac{1}{4}(B^{\prime})^2\right)}%
{\mbox{Det}\left(I-\frac{1}{4}B^2\right)}
\label{lneven}
\end{eqnarray}
using (\ref{lntr}).
To find the eigenvalues of $B'$, $\mbox{Det}(B'-\lambda I)$ can be
expanded
about the middle row to see that the characteristic polynomial of $B'$ 
is proportional to $\lambda$ times the product of the (equal)
characteristic polynomials of the upper-left and lower-right 
$(m{-}4)/2\times (m{-}4)/2$ submatrices of $B'$, which we denote
$B'_1$ and $B'_2$. These submatrices can
be fully diagonalised using the $(m{-}4)/2$ eigenvectors $\psi_k$
of $B$ with $k$ odd: from (\ref{Jparity}), these eigenvectors satisfy
$J\psi_k=-\psi_k$. Hence their middle
components are zero, while the neighbouring two are the
negatives of each other. Given the definition (\ref{BPdef}) of $B'$
this means that projecting each $\psi_k$ for $k$ odd
onto its first $(m{-}4)/2$
components yields $(m{-}4)/2$ independent eigenvectors of $B'_1$
with eigenvalues $\lambda_k$, and likewise for $B'_2$. 
Hence the eigenvalues of $B'$ are
\beq
\lambda_k=2\cos\bigl(\frac{\pi k}{m}\bigr)\,,\qquad k=3,5\dots m{-}3\,,
\label{elist}
\eeq
each with multiplicity $2$, together with $0$.
If $m=2$ mod $4$, then $0$ is also in
the set (\ref{elist}), and so
the algebraic multiplicity of the zero eigenvalue is $3$, even though its 
geometric multiplicity turns out to be only $2$. However, this lack of full 
diagonalisability makes no difference to the computations of traces.

We can now calculate $\ln \goB$. Evaluating
(\ref{lneven}) using the eigenvalues just obtained,
\beqa
 \frac{\mbox{Det}\left(I-\frac{1}{4}(B^{\prime})^2\right)}%
{\mbox{Det}\left(I-\frac{1}{4}B^2\right)}&=&
\frac{\sin^4\frac{3\pi}{m}\sin^4\frac{5\pi}{m}\cdots
\sin^4\frac{(m-3)\pi}{m}}{\sin^2\frac{2\pi}{m}\sin^2
\frac{3\pi}{m}\cdots\sin^2\frac{(m-2)\pi}{m}}\nn\\
&=&\frac{\sin^2\frac{3\pi}{m}\sin^2\frac{5\pi}{m}\cdots
\sin^2\frac{(m-3)\pi}{m}}{\sin^2\frac{2\pi}{m}\sin^2
\frac{4\pi}{m}\cdots\sin^2\frac{(m-2)\pi}{m}}\,.
\label{frac1}
\eeqa
Labelling the final numerator $X$ and the denominator $Y$,
(\ref{trig}) at $n=m$ and $n=m/2$ respectively implies that
\beq
XY=\frac{1}{\sin^4\frac{\pi}{m}}\left(\frac{m}{2^{m-1}}\right)^2~,\qquad
Y=\frac{m^2}{2^{m}}\,.
\eeq
Hence
\beq 
\frac{X}{Y}=\frac{\mbox{Det}\left(I-\frac{1}{4}(B^{\prime})^2\right)}%
{\mbox{Det}\left(I-\frac{1}{4}B^2\right)}=
\left(\frac{2}{m\sin^2\frac{\pi}{m}}\right)^2
\eeq
\nobreak
and (\ref{lneven}) agrees with (\ref{g0beven}). 
\goodbreak

\bigskip
\bigskip

\hrule

\bigskip

\end{document}